\documentclass[english,eqsecnum,nofootinbib,superscriptaddress]{revtex4-2}
\usepackage[T1]{fontenc}
\usepackage[latin9]{inputenc}
\usepackage{color}
\usepackage{babel}
\usepackage{float}
\usepackage{mathtools}
\usepackage{bm}
\usepackage{amsmath}
\usepackage{amssymb}
\usepackage{graphicx}
\usepackage[a4paper]{geometry}
\usepackage[pdfusetitle,
 bookmarks=true,bookmarksnumbered=false,bookmarksopen=false,
 breaklinks=false,pdfborder={0 0 1},backref=false,colorlinks=true]
 {hyperref}

\makeatletter
\usepackage{datetime2}

\makeatother

\begin{document}
\title{Scalar-induced gravitational waves in spatially covariant gravity}
\author{Jiehao Jiang}
\affiliation{School of Physics, Sun Yat-sen University, Guangzhou 510275, China}
\author{Jieming Lin}
\affiliation{Abdus Salam Centre for Theoretical Physics, Imperial College London,
Prince Consort Road, London, SW7 2AZ, UK}
\author{Xian Gao}
\email[Corresponding author: ]{gaoxian@mail.sysu.edu.cn}

\affiliation{School of Physics, Sun Yat-sen University, Guangzhou 510275, China}
\begin{abstract}
We investigate scalar-induced gravitational waves (SIGWs) in the framework
of spatially covariant gravity (SCG), a broad class of Lorentz-violating
modified gravity theories respecting only spatial diffeomorphism invariance.
Extending earlier SCG formulations, we compute the general kernel
function for SIGWs on a flat Friedmann-Lemaître-Robertson-Walker background,
focusing on polynomial-type SCG Lagrangians up to $d=3$, where $d$
denotes the total number of derivatives in each monomial. We derive
explicit expressions for the kernel in the case of power-law time
evolution of the coefficients, and restrict attention to the subset
of SCG operators whose tensor modes propagate at the speed of light,
thereby avoiding late-time divergences in the fractional energy density
of SIGWs. Instead of the usual Newtonian gauge, the breaking of time
reparametrization symmetry in SCG necessitates a unitary gauge analysis.
We compute the energy density of SIGWs for representative parameter
combinations, finding distinctive deviations from general relativity
(GR), including scale-dependent modifications to both the amplitude
and the spectral shape. Our results highlight the potential of stochastic
GW background measurements to probe spatially covariant gravity and
other Lorentz-violating extensions of GR.
\end{abstract}
\maketitle

\section{Introduction}

Einstein's general relativity (GR) remains the standard framework
for describing gravitational phenomena from the solar system to cosmological
scales, having passed numerous observational tests. Nevertheless,
persistent open problems motivate exploration beyond GR. From an observational
perspective, the accelerated expansion of the universe \citep{SupernovaCosmologyProject:1998vns,SupernovaSearchTeam:1998fmf}
suggests the existence of dark energy. Tension between early and late-time
measurements of the Hubble constant \citep{Verde:2019ivm,DiValentino:2021izs}
also suggests the need for new physics. Modified gravity theories
offer a compelling route to address these puzzles (see e.g. \citep{Joyce:2014kja}
for a review). 

The detection of gravitational waves (GWs) by LIGO and Virgo \citep{LIGOScientific:2016fpe,LIGOScientific:2016sjg,LIGOScientific:2016aoc,LIGOScientific:2017vox,LIGOScientific:2017vwq,LIGOScientific:2017ycc,LIGOScientific:2017bnn,LIGOScientific:2018mvr,LIGOScientific:2020zkf,LIGOScientific:2020aai,LIGOScientific:2020stg}
has provided a novel tool for testing gravity on cosmological scales.
GWs are conventionally studied at first order as linear tensor modes
generated during inflation. However, even in the absence of a large
primordial tensor signal, nonlinear mode couplings ensure that scalar
fluctuations inevitably generate second- or higher-order GWs, known
as scalar-induced gravitational waves (SIGWs) \citep{Matarrese:1992rp,Matarrese:1993zf,Ananda:2006af,Baumann:2007zm}.
Recent observations from pulsar timing arrays (PTAs), including NANOGrav
\citep{DeLuca:2020agl,Vaskonen:2020lbd,Kohri:2020qqd,Domenech:2020ers,NANOGrav:2023gor,NANOGrav:2023hde},
PPTA \citep{Reardon:2023gzh}, EPTA \citep{EPTA:2023fyk}, and CPTA
\citep{Xu:2023wog}, have reported a common-spectrum stochastic signal
in the nanohertz band that can be interpreted in terms of SIGWs \citep{Cai:2023dls,Wang:2023ost,Yi:2023mbm,Yi:2023tdk,Harigaya:2023pmw,Liu:2023ymk,Chen:2024twp,Liu:2023hpw}.
Forthcoming space-based interferometers such as LISA \citep{LISA:2017pwj},
DECIGO \citep{Kawamura:2020pcg}, Taiji \citep{Hu:2017mde} and TianQin
\citep{TianQin:2015yph} will further expand sensitivity to SIGWs
across a wide frequency range. 

Because SIGWs are generated during horizon re-entry of enhanced scalar
perturbations, they can serve as a powerful probe of both the small-scale
primordial power spectrum and modifications to the underlying gravitational
theory \citep{Fumagalli:2020nvq,Fu:2019vqc,Choudhury:2024one,Bhattacharya:2023ysp,Choudhury:2023hfm,Gu:2023mmd,Zhao:2023xnh,Fu:2022ssq,Aldabergenov:2022rfc,Arya:2022xzc,Balaji:2022rsy,Rezazadeh:2021clf,Cai:2021wzd,Solbi:2021rse,Zhang:2021rqs,Ahmed:2021ucx,Lin:2020goi,Saito:2008jc,Orlofsky:2016vbd,Nakama:2016gzw,Kohri:2018awv,Kuroyanagi:2018csn,Lu:2020diy,Braglia:2020taf,Chang:2022vlv,Romero-Rodriguez:2021aws,Ota:2021fdv,Zhou:2021vcw,Domenech:2021ztg,Inomata:2021zel,Yuan:2021qgz,Sipp:2022kmb,Zhang:2022dgx,Wang:2023sij}.
Moreover, connections to primordial black hole (PBH) formation \citep{Wang:2016ana,Wang:2021djr,Yi:2022ymw,Papanikolaou:2021uhe,Chen:2021nxo,Kozaczuk:2021wcl,Yi:2022anu,Cang:2022jyc,Zhao:2022kvz,Papanikolaou:2022hkg,Chang:2022nzu,Ghoshal:2023wri,Balaji:2023ehk}
have also heightened the importance of SIGWs. SIGWs depend on nonlinear
mode coupling and thus can probe interaction terms that leave linear
dynamics unchanged. In modified gravity theories, both the evolution
of scalar and tensor perturbations themselves and the coupling between
the scalar and tensor modes can differ from those in GR, resulting
in different SIGW predictions. Analyses have been performed for Brans-Dicke
theory \citep{Yi:2022anu}, $f(R)$ gravity \citep{Zhou:2024doz,Lopez:2025gfu},
Gauss-Bonnet models \citep{Zhang:2021rqs}, Horndeski theory \citep{Domenech:2024drm},
and modified teleparallel gravity \citep{Tzerefos:2023mpe}. In particular,
parity-violating (PV) gravity theories are of special interest, which
predict possible circular polarization in the SIGW background. The
SIGWs in PV gravity have been studied extensively \citep{Zhang:2022xmm,Feng:2023veu,Zhang:2023scq,Zhang:2024vfw,Feng:2024yic,Garcia-Saenz:2023zue,Zhang:2025mps}.
See \citep{Kugarajh:2025rbt} for a recent review and more references
therein.

In this work, we focus on SIGWs in spatially covariant gravity (SCG),
a general class of modified gravity theories respecting only the spatial
covariance \citep{Gao:2014soa,Gao:2014fra}. In this sense, effective
field theory of inflation \citep{Creminelli:2006xe,Cheung:2007st}
and dark energy \citep{Gubitosi:2012hu} as well as the Ho\v{r}ava
gravity \citep{Horava:2009uw,Blas:2009qj} can be viewed as special
cases of SCG. One important advantage of SCG is that due to the separation
between space and time derivatives, it is feasible to build theories
with the desired degrees of freedom (DOFs). In its simplest form without
higher time derivatives, SCG propagates 3 DOFs (two tensor and one
scalar), and thus has a natural correspondence to the single-field
scalar-tensor theories \citep{Gao:2020juc,Gao:2020yzr,Gao:2020qxy,Hu:2021bbo,Hu:2024hzo}.

Cosmological perturbations and linear GWs in SCG theories have been
extensively studied \citep{Fujita:2015ymn,Gao:2019liu,Zhu:2022dfq,Li:2024fxy}.
See also \citep{Gong:2021jgg,Zhu:2022uoq,Zhu:2023rrx,Guo:2025bxz,Wang:2025fhw,Zhang:2024rel,Zhang:2025kcw,Allahyari:2025sbt}
for constraints on Lorentz violation from GWs. Since SCG generically
breaks Lorentz invariance, it allows a richer set of operators than
covariant theories, many of which remain viable after GW170817 constraints
\citep{Gao:2019liu}. In particular, SCG modifies both the scalar
and tensor perturbation sectors, altering the cubic interactions that
source SIGWs. Unlike in GR, the kernel function in SCG can depend
explicitly on the lapse and extrinsic curvature operators, introducing
new time-dependent couplings. 

In this work, we focus on a class of polynomial-type SCG Lagrangians
up to $d=3$, where $d$ is the total number of derivatives in each
monomial. We follow the standard second-order cosmological perturbation
theory to derive the general form of the equations of motion for SIGWs
in SCG. Since there is no time reparametrization symmetry in SCG,
we are not allowed to work in the Newtonian gauge that is commonly
employed in the investigation of SIGWs. Instead, since SCG can be
equivalently viewed as the gauge-fixed version of a scalar-tensor
theory, we work in the so-called unitary gauge. 

Using Green's function methods, we derive the general kernel function
governing SIGW production, with explicit dependence on SCG operators.
Assuming power-law time dependence of the coefficients in the Lagrangian,
we can obtain explicit solutions for the kernel function. We then
concentrate on the contributions from  SCG to the SIGWs during the
radiation-dominated era. To be specific, we will compute the energy
density $\varOmega_{\mathrm{GW}}$ of SIGWs for SCG operators with
luminal tensor modes, and explore parameter combinations illustrating
deviations from GR. As we will see, the SCG produces distinctive deviations
from GR in the fractional energy density of SIGWs.

This paper is organized as follows. In Sec. \ref{sec:scg}, we briefly
review the SCG framework and specify the model we consider. In Sec.
\ref{sec:linear}, we introduce the cosmological perturbations and
derive the equations of motion for the background evolution as well
as the linear scalar and tensor perturbations. In Sec. \ref{sec:cubeom},
we compute the cubic action involving one tensor and two scalar perturbations,
and derive the equations of motion for the SIGWs and in particular
the source term and the kernel function. In Sec. \ref{sec:spec},
we evaluate the energy density of SIGWs for representative parameter
choices, considering both monochromatic and log-normal primordial
scalar spectra, and highlight key deviations from GR. We summarize
and discuss implications in Sec. \ref{sec:con}.

\section{Spatially covariant gravity \label{sec:scg}}

Spatially covariant gravity (SCG) is defined as a class of gravity
theories respecting only the spatial covariance. The general action
of SCG takes the form\footnote{By definition, the SCG action (\ref{scgact}) describes a single foliation
of spacetime. It is possible to generalize it to the case with multiple
foliations, which corresponds to the case with multiple scalar fields
\citep{Yu:2024sed}. Moreover, here we consider only the Riemannian
geometry, in which the metric variables are the only variables to
build the theory. The SCG can be generalized to the case with an independent
connection \citep{Yu:2024drx}.}
\begin{equation}
S=\int\mathrm{d}t\mathrm{d}^{3}xN\sqrt{h}\mathcal{L}\left(t,N,h_{ij},K_{ij},R_{ij},\nabla_{i},\pounds_{\bm{n}},\epsilon_{ijk}\right).\label{scgact}
\end{equation}
Here $N=N\left(t,\bm{x}\right)$ and $h_{ij}=h_{ij}\left(t,\bm{x}\right)$
are the lapse function and 3-dimensional spatial metric in the usual
3+1 or Arnowitt-Deser-Misner (ADM) formalism. They are related to
the 4-dimensional spacetime metric through
\begin{equation}
\text{d}s^{2}=-N^{2}\text{d}t^{2}+h_{ij}(\text{d}x^{i}+N^{i}\text{d}t)(\text{d}x^{j}+N^{j}\text{d}t),
\end{equation}
or in terms of components
\begin{equation}
g_{\mu\nu}=\left(\begin{array}{cc}
g_{00} & g_{0j}\\
g_{i0} & g_{ij}
\end{array}\right)=\left(\begin{array}{cc}
-N^{2}+N^{i}N_{i} & N_{j}\\
N_{i} & h_{ij}
\end{array}\right),
\end{equation}
where $N^{i}$ is the shift vector. In (\ref{scgact}), $K_{ij}$
is the extrinsic curvature defined by
\begin{equation}
K_{ij}\coloneqq\frac{1}{2}\pounds_{\bm{n}}h_{ij}=\frac{1}{2N}\left(\partial_{t}h_{ij}-\pounds_{\vec{N}}h_{ij}\right),
\end{equation}
where $\pounds_{\bm{n}}$ is the Lie derivative with respect to the
normal vector (as a 4-dimensional vector) $n_{\mu}=-N\delta_{\mu}^{0}$
to the spacelike hypersurfaces. The 3-dimensional Ricci tensor $R_{ij}$
and 3-dimensional covariant derivative $\nabla_{i}$ are naturally
included in SCG. In addition, the spatial Levi-Civita tensor $\epsilon_{ijk}$
(with $\epsilon_{123}=\sqrt{h}$) is also allowed in general, which
will introduce parity-violating effects.

The action (\ref{scgact}) is too general to make definitive calculations
for the cosmological perturbations and in particular the scalar-induced
gravitational waves. In the rest of this paper, we concentrate on
a more concrete model of SCG, of which the Lagrangian is a polynomial
built of $K_{ij}$, $R_{ij}$ and their spatial derivatives (i.e.,
without higher temporal derivatives). To be precise, the action is
given by \citep{Gao:2014fra,Fujita:2015ymn}
\begin{equation}
S=\int\mathrm{d}t\mathrm{d}^{3}xN\sqrt{h}\mathcal{L},\label{scgpoly}
\end{equation}
with
\begin{equation}
\mathcal{L}=\mathcal{L}^{(0)}+\mathcal{L}^{(1)}+\mathcal{L}^{(2)}+\mathcal{L}^{(3)}+\mathcal{L}^{(4)},\label{Lag}
\end{equation}
where
\begin{equation}
\mathcal{L}^{(0)}=c_{1}^{(0)},
\end{equation}
\begin{equation}
\mathcal{L}^{(1)}=c_{1}^{(1)}K,
\end{equation}
\begin{equation}
\mathcal{L}^{(2)}=c_{1}^{(2)}K_{ij}K^{ij}+c_{2}^{(2)}K^{2}+c_{3}^{(2)}R,
\end{equation}
\begin{equation}
\mathcal{L}^{(3)}=c_{1}^{(3)}K_{ij}K^{jk}K_{k}^{i}+c_{2}^{(3)}KK_{ij}K^{ij}+c_{3}^{(3)}K^{3}+c_{4}^{(3)}R^{ij}K_{ij}+c_{5}^{(3)}RK,
\end{equation}
and
\begin{align}
\mathcal{L}^{(4)}=\; & c_{1}^{(4)}K_{ij}K^{jk}K_{k}^{i}K+c_{2}^{(4)}(K_{ij}K^{ij})^{2}+c_{3}^{(4)}K_{ij}K^{ij}K^{2}+c_{4}^{(4)}K^{4}\nonumber \\
 & +c_{5}^{(4)}\nabla_{k}K_{ij}\nabla^{k}K^{ij}+c_{6}^{(4)}\nabla_{i}K_{ij}\nabla_{k}K_{j}^{k}+c_{7}^{(4)}\nabla_{i}K^{ij}\nabla_{j}K+c_{8}^{(4)}\nabla_{i}K\nabla^{i}K\nonumber \\
 & +c_{9}^{(4)}R_{ij}K_{k}^{i}K^{jk}+c_{10}^{(4)}R_{ij}K^{ij}K+c_{11}^{(4)}RK_{ij}K^{ij}+c_{12}^{(4)}RK^{2}\nonumber \\
 & +c_{13}^{(4)}R_{ij}R^{ij}+c_{14}^{(4)}R^{2}.
\end{align}
Throughout this paper, $R_{ij}$ and $R$ stand for 3-dimensional
Ricci tensor and scalar, respectively. We classify monomials by numbers
of derivatives, in which $n$ in $\mathcal{L}^{(n)}$ counts the total
number of derivatives in each monomial. See \citep{Hu:2021yaq} for
a detailed classification for more general SCG monomials. In the above,
coefficients $c_{m}^{(n)}$ can be general functions of $t$ and $N$.
Note that $\mathcal{L}^{(2)}$ has included the same monomials that
appear in the 3+1 decomposition of GR but with general coefficients
(as functions of $t$). 

Here we exclude the spatial Levi-Civita tensor $\epsilon_{ijk}$ which
leads to parity violation. The SCG with parity violation and in particular
the polarized linear GWs are investigated in \citep{Gao:2019liu}.
See also \citep{Hu:2014hra} for a recent discussion on more general
SCG monomials with parity violation and their correspondence to the
scalar-tensor theory. We also exclude the spatial derivative of the
lapse function, i.e., the acceleration $a_{i}=\nabla_{i}\ln N$.

Although we have focused on the concrete model (\ref{scgpoly}) with
a polynomial-type Lagrangian, it is still too general and involved
to calculate the cosmological perturbations. To get a better understanding
of features of SIGW in SCG, in the following we make a couple of assumptions
on the general action (\ref{scgpoly}). 

First, we assume $c_{1}^{(0)}=c_{1}^{(0)}(t,N)$ while all other coefficients
are functions of $t$ only for simplicity. Second, due to the large
number of monomials in $\mathcal{L}^{(4)}$, in the following we consider
the Lagrangian up to $\mathcal{L}^{(3)}$, which actually has already
shown characteristic features of SIGWs from SCG. Third, we assume
that the SCG deviates from GR slightly, i.e., all the coefficients
in $\mathcal{L}^{(1)}$ and $\mathcal{L}^{(3)}$ are treated as small
parameters while for coefficients in $\mathcal{L}^{(2)}$, we assume
$c_{i}^{(2)}-c_{i,\mathrm{GR}}^{(2)}$ are small with $c_{1,\mathrm{GR}}^{(2)}=-c_{2,\mathrm{GR}}^{(2)}=c_{3,\mathrm{GR}}^{(2)}=\frac{1}{2}$.
We emphasize that this ``small-deviation'' assumption is both physically
and technically motivated. Physically, modified gravity theories are
already constrained by cosmological and gravitational-wave observations,
so it is natural to explore the vicinity of GR as a conservative baseline.
Technically, treating $c_{1}^{(1)}$, $c_{i}^{(3)}$ and $c_{i}^{(2)}-c_{i,\mathrm{GR}}^{(2)}$
as small ensures that the background and linear dynamics remain close
to GR, so that the radiation-era transfer function can be used at
leading order, while allowing characteristic SCG interactions to appear
already in the kernel through the cubic terms. In addition to this
``small-deviation'' assumption, our parameter space is constrained
by the existence of a radiation-dominated background, and the requirement
of a healthy tensor sector together with observational consistency
of GW propagation speed.

In order to get the solution of SIGWs with general equation of state
parameter $w$ and the speed of sound of the scalar perturbation $c_{s}^{2}$,
we introduce a $k$-essence-like term as the matter field to make
sure the existence of a radiation-dominated period. In a generally
covariant language, the action of $k$-essence is given by
\begin{equation}
\tilde{S}_{k}=\int\mathrm{d}t\mathrm{d}^{3}xN\sqrt{h}\mathcal{L}_{k}\left(\phi,X\right),
\end{equation}
with $X=-(\partial\phi)^{2}/2$. In the SCG language (i.e., when written
in the so-called unitary gauge with $\phi=t$), we have $X=1/2N$.
Therefore, we assume that $\mathcal{L}^{(0)}$ takes the form
\begin{equation}
\mathcal{\tilde{L}}^{(0)}=c_{1}^{(0)}(t,N)\equiv\mathcal{L}_{k}\left(\phi,X\right),
\end{equation}
where $\phi$ and $X$ are understood as being replaced by $t$ and
$1/2N$, respectively. This explains the necessity of assuming $c_{1}^{(0)}$
to be a general function of both $t$ and $N$. Note that if we consider
only the canonical kinetic term for the scalar field, we can get an
arbitrary equation of state parameter $w$ while the speed of sound
$c_{s}^{2}$ is fixed to 1, because a scalar field with canonical
kinetic term always leads to $\delta P=\delta\rho$ in the unitary
gauge. 

With the above discussion on the coefficients, the comparison with
GR is rather straightforward. While the Horndeski theory can also
be written in the form of (\ref{Lag}), but with the 10 coefficients
$c_{m}^{(i)}$'s fully determined by 4 independent coefficients $G_{2},\cdots,G_{5}$
defined in the Horndeski theory \citep{Gleyzes:2013ooa,Fujita:2015ymn}.

\section{Background evolution and linear perturbations \label{sec:linear}}

\subsection{The cosmological perturbations}

When considering perturbations around a Friedmann-Lemaître-Robertson-Walker
(FLRW) background, we parametrize the ADM variables as
\begin{align}
N & =\bar{N}e^{A},\label{pert_N}\\
N_{i} & =\bar{N}a\partial_{i}B,\label{pert_Ni}\\
h_{ij} & =a^{2}e^{2\zeta}\left(e^{\bm{\gamma}}\right)_{ij}.\label{pert_hij}
\end{align}
In the above, $\bar{N}=\bar{N}\left(t\right)$ is the background value
of the lapse function, $a=a\left(t\right)$ is the scale factor. $A,B,\zeta$
are scalar perturbations, and $\gamma_{ij}$ is the tensor perturbation.
The Friedmann-Lema\^{i}tre-Robertson-Walker background thus corresponds
to
\begin{equation}
A=B=\zeta=0,\quad\gamma_{ij}=0.
\end{equation}
Even though one can always set the background value $\bar{N}(t)=1$
in a general covariant theory, we cannot do it here since the SCG
theory has no time-reparametrization symmetry \citep{Fujita:2015ymn,Hu:2021yaq}.
For our purposes, we suppress the vector perturbations as they are
non-dynamical and decay as $a^{-2}$. 

The matrix exponential $e^{\bm{\gamma}}$ is defined as
\begin{equation}
\left(e^{\bm{\gamma}}\right)_{ij}=\delta_{ij}+\gamma_{ij}+\frac{1}{2}\gamma_{i}^{k}\gamma_{kj}+\frac{1}{6}\gamma_{i}^{k}\gamma_{kl}\gamma_{j}^{l}+\cdots,
\end{equation}
where and in the following, the spatial indices in the perturbation
theory are raised and lowered by $\delta^{ij}$ and $\delta_{ij}$,
e.g., $\gamma_{i}^{k}:=\gamma_{ij}\delta^{jk}$. As usual, $\gamma_{ij}$
is defined as a traceless and transverse symmetric tensor,
\begin{equation}
\partial^{i}\gamma_{ij}=0,\qquad\delta^{ij}\gamma_{ij}=0,
\end{equation}
which just corresponds to the tensor perturbation. For simplicity,
we also adopt the notation:
\begin{equation}
\partial^{i}\coloneqq\delta^{ij}\partial_{j},\quad\partial^{2}\coloneqq\partial^{i}\partial_{i}.
\end{equation}

The inverse spatial metric is given by
\begin{align}
h^{ij} & =a^{-2}e^{-2\zeta}\left(e^{-\bm{\gamma}}\right)^{ij}\nonumber \\
 & =a^{-2}e^{-2\zeta}\left(\delta^{ij}-\gamma^{ij}+\frac{1}{2}\gamma_{k}^{i}\gamma^{kj}-\frac{1}{6}\gamma_{k}^{i}\gamma_{l}^{k}\gamma^{lj}+\dots\right).
\end{align}
It is straightforward to check that
\begin{equation}
h^{ij}h_{jk}=\delta_{k}^{i}.
\end{equation}

By using (\ref{pert_N}), we expand $c_{1}^{(0)}(t,N)$ as 
\begin{align}
c_{1}^{(0)}(N,t) & =c_{1,0}^{(0)}+c_{1,0}^{(0)}{}'\left(e^{A}-1\right)+\frac{1}{2}c_{1,0}^{(0)}{}''\left(e^{A}-1\right)^{2},
\end{align}
where and throughout this paper, we follow the notation of \citep{Fujita:2015ymn}
by denoting
\begin{align}
c_{1,0}^{(0)}{}' & =\bar{N}\left.\frac{\partial c_{1}^{(0)}}{\partial N}\right|_{N=\bar{N}},\label{coeff_d1}\\
c_{1,0}^{(0)}{}'' & =\bar{N}^{2}\left.\frac{\partial^{2}c_{1}^{(0)}}{\partial N^{2}}\right|_{N=\bar{N}}.\label{coeff_d2}
\end{align}
At this point, recall that among all the coefficients only $c_{1}^{(0)}$
depends on $N$.

In terms of $\mathcal{L}_{k}\left(\phi,X\right)$ and after fixing
the unitary gauge $\phi=t$, we get the explicit expressions 
\begin{align}
c_{1,0}^{(0)} & =\mathcal{L}_{k}|_{N=\bar{N}},\\
c_{1,0}^{(0)}{}' & =-\bar{N}\left.\frac{\mathcal{L}_{k,X}}{N^{3}}\right|_{N=\bar{N}}=-\frac{\mathcal{L}_{k,X}}{\bar{N}^{2}},\\
c_{1,0}^{(0)}{}'' & =\bar{N}^{2}\left.\left(\mathcal{L}_{k,XX}\frac{1}{N^{6}}+\mathcal{L}_{k,X}\frac{3}{N^{4}}\right)\right|_{N=\bar{N}}=\mathcal{L}_{k,XX}\frac{1}{\bar{N}^{4}}+\mathcal{L}_{k,X}\frac{3}{\bar{N}^{2}}.
\end{align}
Here $\mathcal{L}_{k,X}\equiv\partial\mathcal{L}_{k}/\partial X$
and $\mathcal{L}_{k,XX}\equiv\partial^{2}\mathcal{L}_{k}/\partial X^{2}$.

\subsection{Background evolution}

In the following, we work with the conformal time $\eta$ defined
by $a\mathrm{d}\eta=\bar{N}\mathrm{d}t$ and use a prime to denote
the derivative with respect to the conformal time, e.g., $X'\coloneqq\frac{\mathrm{d}X}{\mathrm{d}\eta}$\footnote{There may be some confusion regarding the notations. Keep in mind
that primes acting on coefficients $c_{m,0}^{(n)}$ denote derivatives
with respect to $N$, as defined in (\ref{coeff_d1}) and (\ref{coeff_d2}).}. The conformal Hubble parameter is defined as usual $\mathcal{H}\coloneqq\frac{1}{a}\frac{\mathrm{d}a}{\mathrm{d}\eta}$. 

By expanding the action (\ref{scgpoly}) around the FLRW background
with respect to the perturbation variables $A$, $B$, $\zeta$ and
$\gamma_{ij}$, we can get the action for the perturbations. The first-order
action for the perturbations takes the form 
\begin{equation}
S_{(1)}=\int\mathrm{d}\eta\mathrm{d}^{3}xa^{4}\left(\mathcal{E}_{A}A+\mathcal{E}_{\zeta}\zeta\right),\label{eq:1orderS}
\end{equation}
where \citep{Fujita:2015ymn}
\begin{equation}
\mathcal{E}_{A}=c_{1,0}^{(0)}+c_{1,0}^{(0)}{}'-3\left(\frac{\mathcal{H}}{a}\right)^{2}\lambda_{1}-6\left(\frac{\mathcal{H}}{a}\right)^{3}\lambda_{2},
\end{equation}
and
\[
\mathcal{E}_{\zeta}=3c_{1,0}^{(0)}+9\frac{\mathcal{H}}{a}c_{1}^{(1)}+9\left(\frac{\mathcal{H}}{a}\right)^{2}\lambda_{1}+9\left(\frac{\mathcal{H}}{a}\right)^{3}\lambda_{2}-\frac{3}{a^{4}}\partial_{\eta}\left[a^{3}\left(c_{1}^{(1)}+2\frac{\mathcal{H}}{a}\lambda_{1}+3\left(\frac{\mathcal{H}}{a}\right)^{2}\lambda_{2}\right)\right].
\]
with
\begin{align}
\lambda_{1} & =c_{1}^{(2)}+3c_{2}^{(2)},\\
\lambda_{2} & =c_{1}^{(3)}+3c_{2}^{(3)}+9c_{3}^{(3)}.
\end{align}

Requiring the vanishing of the first-order action of perturbations
(\ref{eq:1orderS}) leads to the background equations of motion, which
can be written as
\begin{align}
3\mathcal{H}^{2} & =a^{2}\bar{\rho}_{\text{eff}},\label{eq:FM1}\\
\mathcal{H}^{2}+2\mathcal{H}' & =-a^{2}\bar{P}_{\text{eff}}.
\end{align}
In the above, the effective energy density and pressure are defined
to be
\begin{equation}
\bar{\rho}_{\mathrm{eff}}=\bar{\rho}_{k}+3\left(\frac{\mathcal{H}}{a}\right)^{2}(1+\lambda_{1})+6\left(\frac{\mathcal{H}}{a}\right)^{3}\lambda_{2},\label{rho_eff}
\end{equation}
and
\begin{align}
\bar{P}_{\mathrm{eff}}= & \bar{P}_{k}+3\frac{\mathcal{H}}{a}c_{1}^{(1)}+3\left(\frac{\mathcal{H}}{a}\right)^{2}(1+\lambda_{1})+3\left(\frac{\mathcal{H}}{a}\right)^{3}\lambda_{2}\\
 & -\frac{1}{a^{4}}\partial_{\eta}\left[a^{3}\left(c_{1}^{(1)}+2\frac{\mathcal{H}}{a}(1+\lambda_{1})+3\left(\frac{\mathcal{H}}{a}\right)^{2}\lambda_{2}\right)\right],\label{P_eff}
\end{align}
where $\bar{\rho}_{k}=-(c_{1,0}^{(0)}+c_{1,0}^{(0)}{}')$ and $\bar{P}_{k}=\mathcal{L}_{k}=c_{1,0}^{(0)}$
are the energy density and pressure of $k$-essence. 

In this work, we are interested in the case $w_{\mathrm{eff}}\equiv\bar{P}_{\mathrm{eff}}/\bar{\rho}_{\mathrm{eff}}=1/3$,
which corresponds to the radiation-dominated era. For simplicity,
we also assume $w_{k}\equiv\bar{P}_{k}/\bar{\rho}_{k}=1/3$, which
imposes a constraint on the coefficient $c_{1,0}^{(0)}$. According
to (\ref{rho_eff}) and (\ref{P_eff}) , this yields a constraint
on the coefficients
\begin{equation}
\frac{3\frac{\mathcal{H}}{a}c_{1}^{(1)}+3\left(\frac{\mathcal{H}}{a}\right)^{2}(1+\lambda_{1})+3\left(\frac{\mathcal{H}}{a}\right)^{3}\lambda_{2}-\frac{1}{a^{4}}\partial_{\eta}\left[a^{3}\left(c_{1}^{(1)}+2\frac{\mathcal{H}}{a}(1+\lambda_{1})+3\left(\frac{\mathcal{H}}{a}\right)^{2}\lambda_{2}\right)\right]}{3\left(\frac{\mathcal{H}}{a}\right)^{2}(1+\lambda_{1})+6\left(\frac{\mathcal{H}}{a}\right)^{3}\lambda_{2}}=\frac{1}{3}.\label{RDcondition}
\end{equation}

During the radiation-dominated era, equation (\ref{RDcondition})
leads to $a=\alpha\eta$ and $\mathcal{H}=\frac{1}{\eta}$. The first
Friedmann equation (\ref{eq:FM1}) provides a strong constraint on
the coefficients of SCG polynomials, which are supposed to be arbitrary
functions of conformal time, by requiring that $\bar{\rho}_{\text{eff}}$
should decay as $\eta^{-4}$. Moreover, when evaluating the source
terms of the induced gravitational waves, we will find that the combinations
of coefficients $c_{j}^{(i)}$'s are rather complicated. This is because
there are terms with time derivatives on tensor modes, which should
be removed by integration by parts with respect to conformal time. 

For simplicity, in the rest of this paper we follow a similar approach
to \citep{Domenech:2024drm} by choosing a power-law ansatz of coefficients,
i.e.,
\begin{equation}
c_{j}^{(i)}(\eta)=\left(\frac{a}{\mathcal{H}}\right)^{i-2}C_{j}^{(i)}.\label{Cij}
\end{equation}
In (\ref{Cij}), $C_{j}^{(i)}$ are constants and $\bar{\rho}_{k}=-(c_{1,0}^{(0)}+c_{1,0}^{(0)}{}')=-\left(\frac{a}{\mathcal{H}}\right)^{-2}(C_{1,0}^{(0)}+C_{1,0}^{(0)}{}')$
decays as $\eta^{-4}$ as expected. In addition, this ansatz is compatible
with GR since $c_{j}^{(2)}(\eta)$'s are constant in (conformal) time.
Precisely, GR corresponds to
\begin{equation}
C_{1}^{(2)}=-C_{2}^{(2)}=C_{3}^{(2)}=\frac{1}{2},\label{CijGR}
\end{equation}
with all other $C_{j}^{(i)}$ vanishing.

\subsection{Linear perturbations}

In the following, we derive the linear equations of motion for the
scalar and tensor perturbations. Since the scalar and tensor perturbations
are decoupled at linear order, we will discuss them separately.

\subsubsection{Tensor perturbations}

The quadratic action for the tensor perturbations $\gamma_{ij}$ takes
the form
\begin{equation}
S_{(2)}^{\gamma}=\int\mathrm{d}\eta\frac{\mathrm{d}^{3}k}{(2\pi)^{3}}\left(\mathcal{G_{\gamma}}(\gamma_{ij}')^{2}-k^{2}\mathcal{W_{\gamma}}\gamma_{ij}^{2}\right),\label{S2ten}
\end{equation}
where \citep{Fujita:2015ymn}
\begin{align}
\mathcal{G_{\gamma}} & =\frac{a^{2}}{4}\left(c_{1}^{(2)}+3\frac{\mathcal{H}}{a}\left(c_{1}^{(3)}+c_{2}^{(3)}\right)\right),\\
\mathcal{W_{\gamma}} & =\frac{a^{2}}{4}\left(c_{3}^{(2)}+\frac{3}{2}\frac{\mathcal{H}}{a}\left(c_{4}^{(3)}+2c_{5}^{(3)}\right)+\frac{1}{2a}\partial_{\eta}c_{4}^{(3)}\right).
\end{align}

After making the mode decomposition (see Appendix \ref{app:decten}),
we can write the quadratic action for the polarization modes as
\begin{align}
S_{(2)}^{\gamma} & =\sum_{\lambda=\times,+}\int\mathrm{d}\eta\frac{\mathrm{d}^{3}k}{(2\pi)^{3}}\left(\mathcal{G_{\gamma}}\gamma'_{\lambda}(\bm{k},t)\gamma'_{\lambda}(-\bm{k},t)-k^{2}\mathcal{W_{\gamma}}\gamma_{\lambda}(\bm{k},t)\gamma_{\lambda}(-\bm{k},t)\right),\label{S2tenpola}
\end{align}
where $\gamma_{\lambda}$ with $\lambda=+,\times$ are the two polarization
modes for the tensor perturbations.

Variation of (\ref{S2ten}) with respect to $\gamma_{\lambda}$ leads
to the equations of motion for the linear gravitational waves
\begin{equation}
\gamma_{\lambda}''+\frac{2}{\eta}\gamma_{\lambda}'+c_{\mathrm{T}}^{2}k^{2}\gamma_{\lambda}=0,\quad\lambda=\times,+,
\end{equation}
with
\begin{equation}
c_{\mathrm{T}}^{2}=\frac{2C_{3}^{(2)}+5C_{4}^{(3)}+6C_{5}^{(3)}}{2C_{1}^{(2)}+6C_{1}^{(3)}+6C_{2}^{(3)}}.\label{eq:GW speed}
\end{equation}
In (\ref{eq:GW speed}), $C_{j}^{(i)}$ are the constants defined
in (\ref{Cij}).

Since the current astrophysical observations have put stringent constraints
on the propagating speed of the gravitational waves \citep{LIGOScientific:2017vwq,LIGOScientific:2017zic},
we treat (\ref{eq:GW speed}) as a constraint on the coefficients
in the Lagrangian. In the following we choose $c_{\mathrm{T}}^{2}=1$
rather than a free parameter and thus the Green's function for the
tensor perturbations is the same as that of GR. 

\subsubsection{Scalar perturbation}

The quadratic action for the scalar perturbations $A$, $B$ and $\zeta$
is \citep{Fujita:2015ymn}
\begin{align}
S_{(2)}^{\mathrm{sca}}= & \int\mathrm{d}\eta\frac{\mathrm{d}^{3}k}{(2\pi)^{3}}a^{4}\bigg(\zeta'_{\bm{k}}\mathcal{C}_{\zeta'\zeta'}\zeta'_{-\bm{k}}+\zeta_{\bm{k}}\mathcal{C}_{\zeta\zeta'}\zeta'_{-\bm{k}}+\zeta_{\bm{k}}\mathcal{C}_{\zeta\zeta}\zeta_{-\bm{k}}+A_{\bm{k}}\mathcal{C}_{A\zeta'}\zeta'_{-\bm{k}}+A_{\bm{k}}\mathcal{C}_{A\zeta}\zeta_{-\bm{k}}\nonumber \\
 & +A_{\bm{k}}\mathcal{C}_{AA}A_{-\bm{k}}+A_{\bm{k}}\mathcal{C}_{AB}B_{-\bm{k}}+B_{\bm{k}}\mathcal{C}_{B\zeta'}\zeta'_{-\bm{k}}+B_{\bm{k}}\mathcal{C}_{B\zeta}\zeta{}_{-\bm{k}}+B_{\bm{k}}\mathcal{C}_{BB}B_{-\bm{k}}\bigg),\label{S2zAB}
\end{align}
where variously defined coefficients $\mathcal{C}_{\zeta'\zeta'}$,
$\mathcal{C}_{\zeta\zeta'}$ etc. can be found in Appendix \ref{app:coeffsca}. 

In (\ref{S2zAB}), $A$ and $B$ acquire no kinetic term and thus
play the role of auxiliary variables. Variation of (\ref{S2zAB})
with respect to $A$ and $B$ yields the constraint equations
\begin{align}
2\mathcal{C}_{AA}A_{\bm{k}}+\mathcal{C}_{AB}B_{\bm{k}} & =-\mathcal{C}_{A\zeta'}\zeta'_{\bm{k}}-\mathcal{C}_{A\zeta}\zeta_{\bm{k}},\\
\mathcal{C}_{AB}A_{\bm{k}}+2\mathcal{C}_{BB}B_{\bm{k}} & =-\mathcal{C}_{B\zeta'}\zeta'_{-\bm{k}}-\mathcal{C}_{B\zeta}\zeta_{\bm{k}},\label{eq:3}
\end{align}
from which we can solve $A$ and $B$ in terms of $\zeta$ as
\begin{align}
A_{\bm{k}} & =\frac{2\mathcal{C}_{BB}\left(-\mathcal{C}_{A\zeta'}\zeta'_{\bm{k}}-\mathcal{C}_{A\zeta}\zeta_{\bm{k}}\right)-\mathcal{C}_{AB}\left(-\mathcal{C}_{B\zeta'}\zeta'_{-\bm{k}}-\mathcal{C}_{B\zeta}\zeta_{\bm{k}}\right)}{4\mathcal{C}_{AA}\mathcal{C}_{BB}-\mathcal{C}_{AB}^{2}},\label{solA}\\
B_{\bm{k}} & =\frac{2\mathcal{C}_{AA}\left(-\mathcal{C}_{B\zeta'}\zeta'_{-\bm{k}}-\mathcal{C}_{B\zeta}\zeta_{\bm{k}}\right)-\mathcal{C}_{AB}\left(-\mathcal{C}_{A\zeta'}\zeta'_{\bm{k}}-\mathcal{C}_{A\zeta}\zeta_{\bm{k}}\right)}{4\mathcal{C}_{AA}\mathcal{C}_{BB}-\mathcal{C}_{AB}^{2}}.\label{solB}
\end{align}

By plugging the solutions (\ref{solA}) and (\ref{solB}) for $A$
and $B$ in $S_{(2)}^{\zeta,A,B}$, we get the effective quadratic
action for the single variable $\zeta$, which is given by
\begin{align}
S_{(2)}^{\zeta} & =\int\mathrm{d}\eta\frac{\mathrm{d}^{3}k}{(2\pi)^{3}}a^{4}\bigg(\zeta'_{\bm{k}}\mathcal{D}_{\zeta'\zeta'}\zeta'_{-\bm{k}}+\zeta_{\bm{k}}\mathcal{D}_{\zeta\zeta'}\zeta'_{-\bm{k}}+\zeta_{\bm{k}}\mathcal{D}_{\zeta\zeta}\zeta_{-\bm{k}}\bigg).\label{S2zeta}
\end{align}
Varying (\ref{S2zeta}) with respect to $\zeta$ leads to the equations
of motion for $\zeta$,
\begin{equation}
\mathcal{D}_{\zeta'\zeta'}\zeta''_{\bm{k}}+(\partial_{\eta}\mathcal{D}_{\zeta'\zeta'})\zeta'_{\bm{k}}+\left(\frac{1}{2}\partial_{\eta}\mathcal{D}_{\zeta'\zeta}-\mathcal{D}_{\zeta\zeta}\right)\zeta_{\bm{k}}=0.\label{eomzeta1}
\end{equation}
The evaluation of coefficients in (\ref{S2zeta}) and (\ref{eomzeta1})
is quite involved. Fortunately, thanks to the power-law ansatz of
the coefficients (\ref{Cij}), the equation of motion for $\zeta$
(\ref{eomzeta1}) reduces to be
\begin{equation}
\zeta_{\bm{k}}''+\frac{2}{\eta}\zeta_{\bm{k}}'+c_{s}^{2}k^{2}\zeta_{\bm{k}}-9c_{s}^{2}\frac{\left(C_{1}^{(1)}+C_{1}^{(2)}+3C_{2}^{(2)}+1\right)}{\eta^{2}}\zeta_{\bm{k}}=0.\label{eomzeta}
\end{equation}
Here we have deliberately separated terms involving $\zeta_{\bm{k}}$
into two parts. 

Since we assume that the deviation of our model from GR is small,
it is a good approximation to use the solution for $\zeta$ of GR
when evaluating the SIGWs. If we solve (\ref{eomzeta}) perturbatively
and write $\zeta_{\bm{k}}=\zeta_{\bm{k}}^{\text{GR}}+\zeta_{\bm{k}}^{\text{corr}}$
with $\zeta_{\bm{k}}^{\text{corr}}$ the correction to the solution
in GR, the contribution from $\zeta_{\bm{k}}^{\text{corr}}$ to the
SIGWs will be of subleading order, since SIGWs are sourced by terms
of order $\zeta^{2}$. Hence, we can drop the contribution of $\zeta_{\bm{k}}^{\text{corr}}$.

In the case of GR (see (\ref{CijGR})), the last term in (\ref{eomzeta})
is identically vanishing. Therefore, (\ref{eomzeta}) reduces to
\begin{equation}
\zeta_{\bm{k}}''+\frac{2}{\eta}\zeta_{\bm{k}}'+c_{s}^{2}k^{2}\zeta_{\bm{k}}=0,\label{eomzetaGR}
\end{equation}
where 
\begin{equation}
c_{s}^{2}=\frac{P_{k}'}{\rho_{k}'}=-\frac{c_{1,0}^{(0)}{}'}{c_{1,0}^{(0)}{}''+2c_{1,0}^{(0)}{}'}
\end{equation}
is the well-known speed of sound of the scalar perturbation in $k$-essence.
In this case, the solution of (\ref{eomzetaGR}) is given by 
\begin{equation}
\zeta_{\bm{k}}(\eta)=\zeta_{\bm{k},0}j_{0}(c_{s}k\eta),\label{zeta_sol}
\end{equation}
where $j_{0}$ is the spherical Bessel function.

\section{Cubic action and equations of motion for SIGWs \label{sec:cubeom}}

\subsection{Cubic action}

In this section, we derive the equation of motion for the tensor perturbations
induced from the scalar perturbations. There are two equivalent approaches.
The traditional approach is to first derive the equations of motion
for the ADM variables and then expand to the second order in scalar
perturbations. The more convenient approach is to expand the original
action up to the cubic order and focus on the part involving one tensor
and two scalar modes. Then by varying the relevant cubic action, one
may get the equations of motion for the tensor modes sourced by the
quadratic scalar modes.

After some manipulations, the cubic action for one tensor and two
scalar perturbations in momentum space is given by
\begin{align}
S_{(3)}=\int\mathrm{d}\eta\int\frac{\mathrm{d}^{3}k}{(2\pi)^{3}}\int\frac{\mathrm{d}^{3}l}{(2\pi)^{3}}a^{4} & \bigg\{\mathcal{F}_{1}\gamma_{ij}(-\bm{k})l^{i}l^{j}A_{\bm{q}}B_{\bm{l}}+\mathcal{F}_{2}\gamma_{ij}(-\bm{k})l^{i}l^{j}B_{\bm{q}}B_{\bm{l}}+\mathcal{F}_{3}\gamma_{ij}(-\bm{k})l^{i}l^{j}A_{\bm{q}}\zeta_{\bm{l}}\nonumber \\
 & +\mathcal{F}_{4}\gamma_{ij}(-\bm{k})l^{i}l^{j}B_{\bm{q}}\zeta_{\bm{l}}+\mathcal{F}_{5}\gamma_{ij}(-\bm{k})l^{i}l^{j}\zeta_{\bm{q}}B_{\bm{l}}+\mathcal{F}_{6}\gamma_{ij}(-\bm{k})l^{i}l^{j}\zeta'_{\bm{q}}B_{\bm{l}}\nonumber \\
 & +\mathcal{F}_{7}\gamma_{ij}(-\bm{k})l^{i}l^{j}\zeta_{\bm{q}}\zeta_{\bm{l}}+\mathcal{F}_{8}\gamma_{ij}(-\bm{k})l^{i}l^{j}\zeta'_{\bm{q}}\zeta_{\bm{l}}\nonumber \\
 & +\mathcal{G}_{1}\gamma'_{ij}(-\bm{k})l^{i}l^{j}A_{\bm{q}}B_{\bm{l}}+\mathcal{G}_{2}\gamma'_{ij}(-\bm{k})l^{i}l^{j}B_{\bm{q}}B_{\bm{l}}+\mathcal{G}_{3}\gamma'_{ij}(-\bm{k})l^{i}l^{j}\zeta'_{\bm{q}}B_{\bm{l}}\nonumber \\
 & +\mathcal{G}_{4}\gamma'_{ij}(-\bm{k})l^{i}l^{j}\zeta_{\bm{q}}B_{\bm{l}}+\mathcal{G}_{5}\gamma'_{ij}(-\bm{k})l^{i}l^{j}\zeta_{\bm{q}}\zeta_{\bm{l}}\bigg\},\label{S3}
\end{align}
where the coefficients $\mathcal{F}_{1}$, $\mathcal{F}_{2}$ etc.
can be found in Appendix \ref{app:coeffS3}, $\bm{k}$, $\bm{q}$
and $\bm{l}$ are momenta of the relevant modes. Note in (\ref{S3}),
$\bm{q}=\bm{k}-\bm{l}$ due to the conservation of momenta.

We can now replace the auxiliary variables $A$ and $B$ in terms
of $\zeta$ in the cubic action (\ref{S3}). Note that up to the cubic
order in perturbations, the first order solutions (\ref{solA}) and
(\ref{solB}) are sufficient. For later convenience, we follow the
convention of \citep{Ananda:2006af,Baumann:2007zm} and define two
dimensionless parameters $v=\frac{q}{k}$ and $u=\frac{|\bm{k}-\bm{q}|}{k}$.
A technical issue arises when simply replacing $A_{\bm{q}}$ and $B_{\bm{q}}$
by their first order solutions, which generally leads to a source
function $f$ that will change under the exchange $u\leftrightarrow v$
if we use integration by parts inappropriately to simplify the cubic
action. To get a standard source function $f$, we need to symmetrize
terms in $S_{(3)}$ and $S_{(3),\text{GR}}$ to be
\begin{align}
 & \int\frac{\mathrm{d}^{3}q}{(2\pi)^{3}}F(\bm{k},\bm{q},|\bm{k}-\bm{q}|,\eta)X_{\bm{q}}Y_{\bm{k}-\bm{q}}\gamma_{\lambda}(-\bm{k})e_{ij}^{\lambda}(\bm{k})q^{i}q^{j}\nonumber \\
=\; & \frac{1}{2}\int\frac{\mathrm{d}^{3}q}{(2\pi)^{3}}\left[F(\bm{k},\bm{q},|\bm{k}-\bm{q}|,\eta)X_{\bm{q}}Y_{\bm{k}-\bm{q}}+F(\bm{k},|\bm{k}-\bm{q}|,\bm{q},\eta)X_{\bm{k}-\bm{q}}Y_{\bm{q}}\right]\gamma_{\lambda}(-\bm{k})e_{ij}^{\lambda}(\bm{k})q^{i}q^{j},
\end{align}
where $X$ and $Y$ stand for the scalar perturbations (i.e., $A$,
$B$ or $\zeta$).

After polarization decomposition, the cubic action in the radiation-dominated
era can be written as
\begin{align}
S_{(3)}^{\zeta\zeta\gamma}= & \sum_{\lambda=+,\times}\int\frac{\mathrm{d}^{3}k}{(2\pi)^{3}}\int\frac{\mathrm{d}^{3}q}{(2\pi)^{3}}\mathrm{d}\eta\,a^{2}(\eta)\gamma_{\lambda}(-\bm{k})e_{ij}^{\lambda}(\bm{k})q^{i}q^{j}\nonumber \\
 & \times\bigg\{\mathcal{T}_{\zeta\zeta}\zeta_{\bm{q}}\zeta_{\bm{k}-\bm{q}}+\mathcal{T}_{\zeta\zeta'}\zeta_{\bm{q}}\zeta'_{\bm{k}-\bm{q}}+\mathcal{T}_{\zeta'\zeta}\zeta'_{\bm{q}}\zeta_{\bm{k}-\bm{q}}+\mathcal{T}_{\zeta'\zeta'}\zeta'_{\bm{q}}\zeta'_{\bm{k}-\bm{q}}\nonumber \\
 & -\frac{1}{a(\eta)^{2}}\partial_{\eta}\left[a^{2}\text{\ensuremath{\left(\mathcal{X}_{\zeta\zeta}\zeta_{\bm{q}}\zeta_{\bm{k}-\bm{q}}+\mathcal{X}_{\zeta\zeta'}\zeta_{\bm{q}}\zeta'_{\bm{k}-\bm{q}}+\mathcal{X}_{\zeta'\zeta}\zeta'_{\bm{q}}\zeta_{\bm{k}-\bm{q}}+\mathcal{X}_{\zeta'\zeta'}\zeta'_{\bm{q}}\zeta'_{\bm{k}-\bm{q}}\right)}}\right]\bigg\}.\label{eq:4}
\end{align}
where we have used $a\propto\eta\mathcal{\text{ and }H}=\frac{1}{\eta}$
for the radiation-dominated era. The explicit expressions of the coefficients
$\mathcal{T}_{\zeta\zeta}$ etc. can be found in Appendix \ref{app:coeffS3}. 

By combining the quadratic action for the tensor modes (\ref{S2tenpola}),
the cubic action leads to the equations of motion for the SIGWs
\begin{equation}
\gamma_{\lambda}''+\frac{2}{\eta}\gamma_{\lambda}'+c_{T}^{2}k^{2}\gamma_{\lambda}=\mathcal{S}(k,\eta),\quad\lambda=+,\times,\label{eomSIGWs}
\end{equation}
where $\mathcal{S}\left(k,\eta\right)$ is the source term quadratic
in the scalar perturbation $\zeta$. Following the steps of \citep{Domenech:2021ztg}
and considering the solution of $\zeta$ in (\ref{zeta_sol}), the
source term can be written as
\begin{equation}
\mathcal{S}(k,\eta)=\int\frac{\text{d}^{3}q}{(2\pi)^{3}}e_{ij}^{\lambda}(\bm{k})q^{i}q^{j}\zeta_{\mathbf{q}}\zeta_{\mathbf{k-q}}f(\eta,q,|\bm{k}-\bm{q}|).\label{source}
\end{equation}
The source function $f$ depends on the concrete model, which is quite
complicated in our case. 

Equation (\ref{eomSIGWs}) can be simplified by introducing $z_{\lambda}=\eta\gamma_{\mathbf{\lambda}}$:
\begin{equation}
z_{\lambda}''+k^{2}z_{\lambda}=\eta\mathcal{S}(k,\eta),\label{eqofz}
\end{equation}
which can be solved by the Green's function method
\begin{equation}
z_{\lambda}=\int_{0}^{\eta}\text{d}\tilde{\eta}\frac{1}{k}\sin(k\eta-k\tilde{\eta})\tilde{\eta}\mathcal{S_{\gamma}}(k,\tilde{\eta}),
\end{equation}
or equivalently
\begin{equation}
\gamma_{\lambda}=\int_{0}^{\eta}\text{d}\tilde{\eta}\frac{1}{k}\sin(k\eta-k\tilde{\eta})\frac{\tilde{\eta}}{\eta}\mathcal{S_{\gamma}}(k,\tilde{\eta}).\label{gamma_gf}
\end{equation}

We follow the convention of \citep{Kohri:2018awv} to calculate the
power spectrum $\mathcal{P}_{\gamma}$ for the tensor modes, which
is defined by 
\begin{equation}
\left\langle \gamma_{\mathbf{\lambda}}(\eta,\boldsymbol{k})\gamma_{\mathbf{\lambda'}}(\eta,\boldsymbol{k}')\right\rangle =\delta_{\lambda\lambda'}\delta^{3}(\boldsymbol{k}+\boldsymbol{k}')\frac{2\pi^{2}}{k^{3}}\mathcal{P}_{\gamma}(\eta,k),
\end{equation}
and the concrete expression of $\mathcal{P}_{\gamma}$ can be obtained
by using the solution (\ref{gamma_gf}) for $\gamma_{\lambda}$, though
the calculation is quite involved. After some simplification, we arrive
at \citep{Kohri:2018awv}
\begin{equation}
\mathcal{P}_{\gamma}(\eta,k)=\frac{1}{4}\int_{0}^{\infty}\mathrm{d}v\int_{|1-v|}^{1+v}\mathrm{d}u\left(\frac{4v^{2}-(1+v^{2}-u^{2})^{2}}{4uv}\right)^{2}I^{2}(v,u,x)\mathcal{P_{\zeta}}(kv)\mathcal{P_{\zeta}}(ku),
\end{equation}
where
\begin{align}
I(v,u,x) & =\int_{0}^{x}\mathrm{d}\bar{x}\frac{a(\bar{\eta})}{a(\eta)}kG_{k}(\eta,\bar{\eta})f(v,u,\bar{x}),
\end{align}
is the kernel function in which the information of the Green's function
and the source is encoded, and $\mathcal{P_{\zeta}}$ is the power
spectrum of scalar perturbation defined by
\begin{equation}
\left\langle \zeta(\eta,\boldsymbol{k})\zeta(\eta,\boldsymbol{k}')\right\rangle =\delta^{3}(\boldsymbol{k}+\boldsymbol{k}')\frac{2\pi^{2}}{k^{3}}\mathcal{P}_{\zeta}(\eta,k).
\end{equation}

In the following, we focus on the case of $c_{s}^{2}=1$ in order
to obtain an explicit expression for the source function $f$ and
a better understanding of the kernel. Later we will numerically evaluate
the case of $c_{s}^{2}=1/3$ and show the power spectrum in figures. 

By using the power-law solutions and $\zeta_{\bm{k}}=\zeta_{\bm{k},0}T_{k}$,
with $T_{k}$ the transfer function of the scalar perturbation given
in (\ref{zeta_sol}), the source function $f$ can be written as
\begin{equation}
f=\frac{\mathcal{M}_{1}}{2}T_{q}T_{|\bm{k}-\bm{q}|}+\mathcal{M}_{2}T'_{q}T_{|\bm{k}-\bm{q}|}+\frac{\mathcal{M}_{3}}{2}T'_{q}T'_{|\bm{k}-\bm{q}|}+\mathcal{M}_{4}T''_{q}T_{|\bm{k}-\bm{q}|}+\mathcal{M}_{5}T''_{q}T'_{|\bm{k}-\bm{q}|},\label{eq:f}
\end{equation}
where
\begin{align}
\mathcal{M}_{1} & =m_{1}^{(0)}+m_{1}^{(2)}\eta^{2},\\
\mathcal{M}_{2} & =m_{2}^{(-1)}\frac{1}{\eta}+m_{2}^{(1)}\eta+m_{2}^{(3)}\eta^{3},\\
\mathcal{M}_{3} & =m_{3}^{(0)}+m_{3}^{(2)}\eta^{2},\\
\mathcal{M}_{4} & =m_{4}^{(0)}+m_{4}^{(2)}\eta^{2},\\
\mathcal{M}_{5} & =m_{5}^{(1)}\eta.
\end{align}
Here $m_{j}^{(i)}$ are functions of momenta and their explicit expressions
are given in Appendix \ref{app:srcfun}. In (\ref{eq:f}), terms like
$m_{2}^{(3)},m_{1}^{(2)},m_{4}^{(2)},m_{3}^{(2)}$ will contribute
to higher orders of $x\equiv k\eta$. Unfortunately, we cannot find
a simple solution where these terms vanish because they are functions
of momenta $\left|\bm{k}-\bm{q}\right|$, $q$ and $k$ in general. 

Thus the source function $f$ can be written in the following form
(with $c_{s}^{2}=1$)
\begin{align}
f=\; & \left(d_{0}^{-}+\frac{d_{2}^{-}}{x^{2}}+\frac{d_{4}^{-}}{x^{4}}\right)\cos(ux-vx)+\left(d_{0}^{+}+\frac{d_{2}^{+}}{x^{2}}+\frac{d_{4}^{+}}{x^{4}}\right)\cos(ux+vx)\nonumber \\
 & +\left(xd_{x}^{-}+\frac{d_{1}^{-}}{x}+\frac{d_{3}^{-}}{x^{3}}\right)\sin(ux-vx)+\left(xd_{x}^{+}+\frac{d_{1}^{+}}{x}+\frac{d_{3}^{+}}{x^{3}}\right)\sin(ux+vx),\label{srcf}
\end{align}
where the expressions for $d_{i}$ are given in Appendix \ref{app:srcfun}.
The source function $f$ shares the same general form as that of \citep{Domenech:2024drm}.
The coefficients satisfy $d_{i}^{-}(u,v)=d_{i}^{+}(u,-v)$ as expected
but contain extra terms that will contribute to higher orders of $x$.
This will lead to an apparent late-time divergence as we will discuss
later.

\subsection{More on the source function and the kernel}

In the following, we analyze the source function and kernel in more
detail. We first isolate the GR contribution in the unitary gauge
as a reference, and then extend the discussion to the full SCG case
with general polynomial operators.

\subsubsection{GR part}

The cubic action (\ref{eq:4}) contains a number of coefficients,
which is complicated to be analyzed. Usually, the SIGWs are evaluated
in the Newtonian gauge. The SCG corresponds to a generally covariant
scalar-tensor theory written in the unitary gauge. In order to get
some important features in the unitary gauge, we first consider the
GR part of the cubic action,
\begin{align}
S_{(3),\mathrm{GR}}^{\zeta\zeta\gamma}=\; & \sum_{\lambda=+,\times}\int\frac{\mathrm{d}^{3}k}{(2\pi)^{3}}\int\frac{\mathrm{d}^{3}q}{(2\pi)^{3}}\mathrm{d}\eta\,a^{2}(\eta)\gamma_{\lambda}(-\bm{k})e_{ij}^{\lambda}(\bm{k})q^{i}q^{j}\nonumber \\
 & \times\Big(\mathcal{T}_{\zeta\zeta}^{(\mathrm{GR})}\zeta_{\bm{q}}\zeta_{\bm{k}-\bm{q}}+\mathcal{T}_{\zeta\zeta'}^{(\mathrm{GR})}\zeta_{\bm{q}}\zeta'_{\bm{k}-\bm{q}}+\mathcal{T}_{\zeta'\zeta}^{(\mathrm{GR})}\zeta'_{\bm{q}}\zeta_{\bm{k}-\bm{q}}+\mathcal{T}_{\zeta'\zeta'}^{(\mathrm{GR})}\zeta'_{\bm{q}}\zeta'_{\bm{k}-\bm{q}}\nonumber \\
 & +\mathcal{T}_{\zeta''\zeta}^{(\mathrm{GR})}\zeta''_{\bm{q}}\zeta_{\bm{k}-\bm{q}}+\mathcal{T}_{\zeta\zeta''}^{(\mathrm{GR})}\zeta_{\bm{q}}\zeta''_{\bm{k}-\bm{q}}+\mathcal{T}_{\zeta''\zeta'}^{(\mathrm{GR})}\zeta''_{\bm{q}}\zeta'_{\bm{k}-\bm{q}}+\mathcal{T}_{\zeta'\zeta''}^{(\mathrm{GR})}\zeta'_{\bm{q}}\zeta''_{\bm{k}-\bm{q}}\Big),\label{S3GR}
\end{align}
where we have used integration by parts with respect to the conformal
time to eliminate time derivatives on tensor modes. The coefficients
$\mathcal{T}^{(\mathrm{GR})}$'s are functions of conformal time and
momenta $\bm{q}$ and $\bm{k}$, which are different from those in
the Newtonian gauge where all the coefficients are simply functions
of conformal time. Note that the last two terms in the third line
of (\ref{S3GR}) contain three temporal derivatives in total. Such
 terms do not appear in Newtonian gauge. 

The GR part of the source function $f$ in the unitary gauge can be
written as
\begin{align}
f_{\mathrm{GR}}=\; & \bigg(-\frac{1}{u^{3}v^{3}x^{4}}+\frac{2}{u^{3}vx^{4}}-\frac{v}{2u^{3}x^{2}}+\frac{1}{2u^{3}vx^{2}}+\frac{2}{uv^{3}x^{4}}\nonumber \\
 & -\frac{u}{2v^{3}x^{2}}+\frac{1}{2uv^{3}x^{2}}-\frac{1}{uvx^{2}}+\frac{u}{4v}+\frac{v}{4u}-\frac{1}{4uv}\bigg)\sin(ux)\sin(vx)\nonumber \\
 & +\left(-\frac{1}{u^{2}v^{2}x^{2}}+\frac{2}{u^{2}x^{2}}+\frac{2}{v^{2}x^{2}}-\frac{1}{2}\right)\cos(ux)\cos(vx)\nonumber \\
 & +\left(\frac{1}{u^{3}v^{2}x^{3}}-\frac{2}{u^{3}x^{3}}-\frac{2}{uv^{2}x^{3}}+\frac{u}{2v^{2}x}-\frac{1}{2uv^{2}x}+\frac{1}{ux}\right)\sin(ux)\cos(vx)\nonumber \\
 & +\left(\frac{1}{u^{2}v^{3}x^{3}}-\frac{2}{u^{2}vx^{3}}+\frac{v}{2u^{2}x}-\frac{1}{2u^{2}vx}-\frac{2}{v^{3}x^{3}}+\frac{1}{vx}\right)\cos(ux)\sin(vx).
\end{align}
Note the expression of $f_{\mathrm{GR}}$ is manifestly symmetric
in $u,v$. 

The GR part of the kernel is
\begin{align}
I_{\mathrm{GR}}(u,v,x,k)=\; & -\frac{1}{8u^{3}v^{3}x^{2}}\bigg\{-x\left(u^{2}+v^{2}-1\right)^{2}\sin x\nonumber \\
 & \times\left[\text{Ci}((u-v+1)x)+\text{Ci}((-u+v+1)x)-\text{Ci}((u+v+1)x)-\text{Ci}((-u-v+1)x)\right]\nonumber \\
 & +x\left(u^{2}+v^{2}-1\right)\sin x\left[\left(u^{2}+v^{2}-1\right)\log\left(\frac{|1-(u+v)^{2}|}{|1-(u-v)^{2}|}\right)-4uv\right]\nonumber \\
 & +2\sin(vx)\left[\left(u^{2}\left(v^{2}x^{2}-4\right)-4v^{2}+2\right)\sin(ux)+2uv^{2}x\cos(ux)\right]+4u^{2}vx\sin(ux)\cos(vx)\nonumber \\
 & +x\left(u^{2}+v^{2}-1\right)^{2}\cos x\nonumber \\
 & \times\left[\text{Si}((u-v+1)x)+\text{Si}((-u+v+1)x)-\text{Si}((u+v+1)x)-\text{Si}(-ux-vx+x)\right]\bigg\}.\label{IGR}
\end{align}
Note that there are terms $\sim\frac{1}{8u^{3}v^{3}x^{2}}\sin(vx)\sin(ux)u^{2}\left(v^{2}x^{2}-4\right)$
in the fourth line of (\ref{IGR}), which diverges when taking a late-time
limit $(x\gg1)$. This apparent divergence does not signal a growth
of the GW signal. In a generally covariant theory such as GR, the
late-time observable is gauge invariant, and the apparently divergent
terms that arise in the unitary gauge cancel in gauge-invariant quantities
\citep{Domenech:2024drm}. Equivalently, one may perform the second-order
gauge transformation to Newtonian gauge, in which the late-time kernel
contains only the standard luminal (finite) contributions. However,
SCG respects only the spatial covariance and time reparametrization
invariance is absent, so we cannot switch to the Newtonian gauge within
the same theory. The unitary gauge is instead the natural description
of the SCG action. Nevertheless, the identification of the physical
SIGW signal remains robust. The observable stochastic background is
carried by the freely propagating tensor modes with propagation speed
$c_{\mathrm{T}}=1$, while the additional pieces correspond to non-luminal
or ``strain'' contributions associated with scalar-tensor mixing.
This motivates the separation of the kernel into luminal (finite)
and non-luminal (potentially divergent) parts in the general SCG case
discussed next.

\subsubsection{SCG with general polynomials}

After integrating the Green's function with the power-law ansatz,
the kernel of SCG theory with general polynomials takes the form
\begin{align}
I(u,v,x,k) & =I_{\mathrm{o}}+I_{\mathrm{p}},\label{eq:kernelscg}
\end{align}
where
\begin{align}
I_{\mathrm{o}}=\; & \frac{\sin x}{x}\mathcal{Y}_{1}(u,v)\left[\text{Ci}((u-v+1)x)+\text{Ci}((-u+v+1)x)-\text{Ci}((u+v+1)x)-\text{Ci}(-(u+v-1)x)\right]\nonumber \\
 & +\frac{\cos x}{x}\mathcal{Y}_{2}(u,v)\left[\text{Si}((u-v+1)x)+\text{Si}((-u+v+1)x)-\text{Si}((u+v+1)x)-\text{Si}(-(u+v-1)x)\right]\nonumber \\
 & +\frac{\sin x}{x}\mathcal{Y}_{3}(u,v)+\frac{\cos x}{x}\mathcal{Y}_{4}(u,v),
\end{align}
and
\begin{align}
I_{\mathrm{p}}=\; & xY_{1}(u,v)\sin(ux-vx)+Y_{2}(u,v)\cos(ux-vx)+\frac{1}{x}Y_{3}(u,v)\sin(ux-vx)+\frac{1}{x^{2}}Y_{4}(u,v)\cos(ux-vx)\nonumber \\
 & +(v\to-v).
\end{align}
In the above, we separate the kernel into two parts. $I_{\mathrm{o}}$
stands for ``ordinary'' terms that lead to a finite power spectrum,
which we regard as the main part of the kernel. On the other hand,
$I_{\mathrm{p}}$ stands for ``particular'' terms that are higher
order in $x$, which may cause late-time divergence and include terms
oscillating as $\sin(ux-vx)$. 

The appearance of the $\sin(ux-vx)$ structure in $I_{\mathrm{p}}$
can be viewed as a violation of Huygens' principle. For the luminal
propagation of a massless field, the retarded Green's function has
support on the light cone, whereas non-luminal propagation generically
allows ``tail'' support inside the light cone. The phase $(u-v)x$
reflects precisely such non-luminal contributions in the effective
description. It is these terms that can produce apparent late-time
growth of $\varOmega_{\mathrm{GW}}$. In this work we are interested
in the stochastic background carried by the freely propagating tensor
modes. Accordingly, we will extract the luminal component and discard
the non-luminal strain contributions. We emphasize that we cannot
directly attribute the presence of these non-luminal contributions
(and thus the intra-cone GW modes that violate the Huygens' principle)
to thee breaking Lorentz symmetry. In fact, such non-luminal oscillatory
terms also appear in GR, though in the Newtonian gauge they are not
present at the leading order \citep{Domenech:2024drm,Ali:2025xvj,Domenech:2025zvi}. 

The analytical expressions of $\mathcal{Y}_{i}$ are very lengthy
and tedious, and we prefer not to present them in the paper. Note
that in the Newtonian gauge, all the terms oscillating in $\sin(ux-vx)$
are excluded at late times because they do not appear at the leading
order of $1/x$. This provides an explicit diagnostic for what we
mean by the luminal SIGW component. At late times, it is the part
scaling as $1/x$ (or decaying faster) and oscillating with the GW
phase $x=k\eta$ (more generally $x=c_{\mathrm{T}}k\eta$), which
is precisely the component computed directly in Newtonian gauge in
covariant theories. The additional $\sin(ux-vx)$ terms collected
in $I_{\mathrm{p}}$ are therefore identified as non-luminal contributions
in unitary gauge, which are the origin of the apparent late-time divergences.

\section{Energy density of SIGWs \label{sec:spec}}

As mentioned above, (\ref{eq:kernelscg}) generically contains pieces
that can yield apparent late-time divergences in $\varOmega_{\mathrm{GW}}$.
These divergences originate from the particular component $I_{\mathrm{p}}$,
whose oscillatory structure $\sin(ux-vx)$ corresponds to non-luminal
contributions from the strain components induced by scalar-tensor
mixing. Following the physical criterion advocated in the literature,
we identify the SIGW signal with the freely propagating tensor modes,
i.e. the luminal component that propagates with the GW speed $c_{\mathrm{T}}$
(here we have chosen $c_{\mathrm{T}}=1$) \citep{Ali:2020sfw}. This
means that we keep the late-time luminal part of the kernel (the $I_{\mathrm{o}}$
contribution scaling as $1/x$ or faster) and discard the $I_{\mathrm{p}}$
pieces responsible for the divergence. Equivalently, for the $\sin(ux-vx)$
phase, only the condition $u-v=1$ contributes to the luminal propagation.
All other $\sin(ux-vx)$ contributions are excluded from $\varOmega_{\mathrm{GW}}$.
With this luminal projection, the remaining kernel decays as $x^{-1}$,
and the fractional energy density approaches a finite late-time limit.

We follow the formalism of \citep{Kohri:2018awv} and write the power
spectrum of tensor perturbations as
\begin{equation}
\mathcal{P}_{\gamma}(\eta,k)=2\int_{0}^{\infty}\text{d}t\int_{-1}^{1}\text{d}s\left[\frac{t(2+t)(s^{2}-1)}{(1-s+t)(1+s+t)}\right]^{2}I^{2}(u,v,x)P_{\zeta}(kv)P_{\zeta}(ku),\label{spec}
\end{equation}
where $u=(t+s+1)/2$ and $v=(t-s+1)/2.$ Since $C_{1}^{(1)}$ is absent
in the kernel, we will use the parameters 
\begin{equation}
\beta_{i}=C_{i}^{(2)}-C_{i,\mathrm{GR}}^{(2)},
\end{equation}
and $C_{i}^{(3)}$ in the following sections, which we all assume
to be small parameters. We thus take $|\beta_{i}|\ll1$ and $|C_{i}^{(3)}|\ll1$
as a controlled expansion around GR. The admissible coefficient space
is not arbitrary. The radiation-era background requires the constraint
(\ref{RDcondition}), and we additionally impose the luminal tensor
condition (\ref{eq:GW speed}). 

In the following, we vary one particular subset of coefficients at
a time, and set others to their GR values or to zero. This will diagnose
which SCG terms dominantly reshape the fractional energy density $\varOmega_{\mathrm{GW}}$
of SIGWs. We will discuss two cases of the primordial spectrum for
the scalar perturbation, one is the monochromatic spectrum, the other
is the log-normal spectrum.

\subsection{Monochromatic primordial spectrum}

We first consider a monochromatic primordial spectrum for the curvature
perturbation in the form of a Dirac delta function
\begin{equation}
P_{\zeta}=A_{\zeta}\delta\left(\log\frac{k}{k_{\ast}}\right),
\end{equation}
which represents an extremely narrow enhancement of scalar power centered
at $k=k_{\ast}$.

In the case of GR, when $c_{s}^{2}=1$, the corresponding analytic
expression for the energy density of SIGWs $\varOmega_{\mathrm{GW}}$
in the unitary gauge is given by \citep{Kohri:2018awv}
\begin{align}
\varOmega_{\mathrm{GR}}=\; & \frac{A_{\zeta}^{2}}{6144}k^{2}\left(k^{4}-6k^{2}+8\right)^{2}(k^{2}-2)\nonumber \\
 & \times\left(\left(k^{2}-2\right)\log\left(\left|\frac{k^{2}-4}{k^{2}}\right|\right)+8\right)\log\left(\left|\frac{k^{2}-4}{k^{2}}\right|\right)\Theta(2-k).
\end{align}
The Heaviside factor arises from momentum conservation as in the Newtonian
gauge. In other words, it enforces the triangle condition $|u-v|\leq1\leq u+v$,
manifesting as a sharp cut-off in $k/k_{\ast}$ space.

In the case of SCG, the explicit expression for the fractional energy
density is rather lengthy. In the following we simply show the numerical
results with different choices of values of the coefficients. We have
a total of 8 parameters (i.e., coefficients in the Lagrangian) in
$\mathcal{L}^{(2)}$ and $\mathcal{L}^{(3)}$, but not all choices
of these parameter values are accessible. In what follows, we focus
on several interesting choices of parameters and show the energy density
of SIGWs.

\subsubsection{Case 1: $\beta_{1}=-\beta_{2}=\beta_{3}=y$}

First we consider the case with $\beta_{1}=-\beta_{2}=\beta_{3}=y$,
where $y$ is constant. This case corresponds to a generally covariant
limit of SCG. 

\begin{figure}[H]
\includegraphics[scale=0.4]{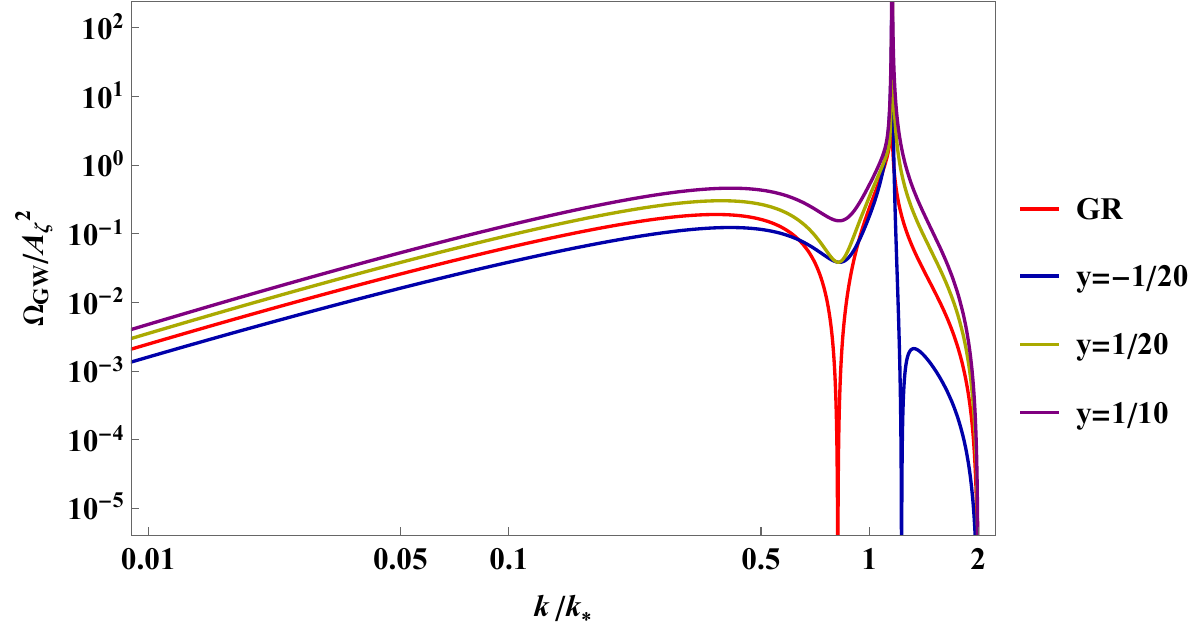}\includegraphics[scale=0.4]{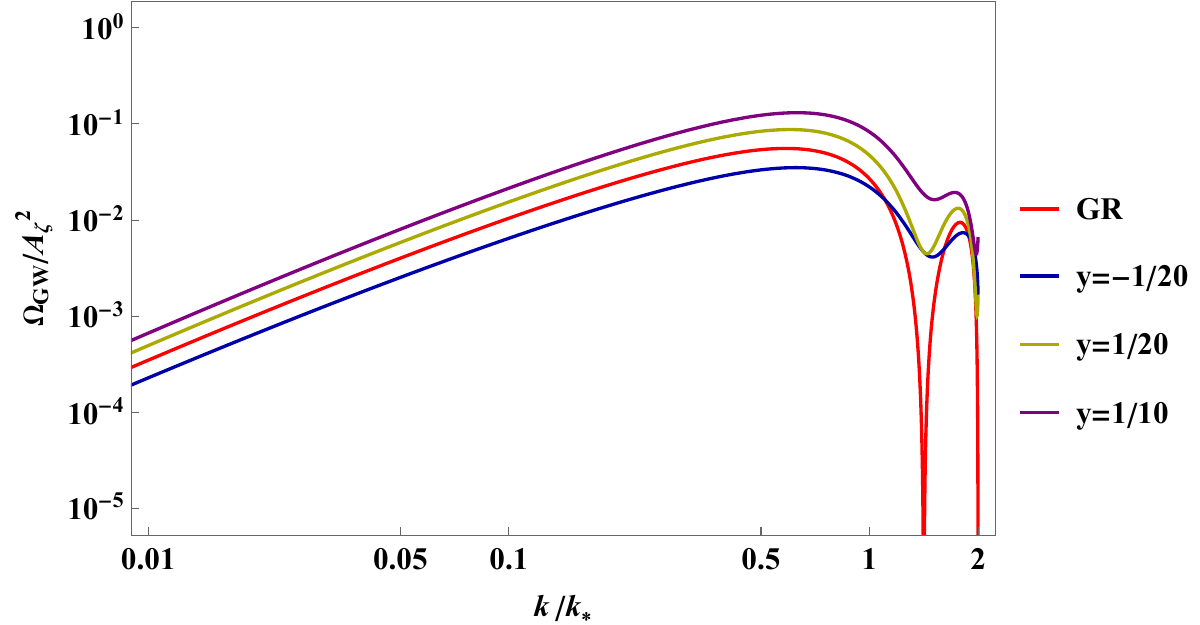}

\caption{Energy density of SIGWs with $\beta_{1}=-\beta_{2}=\beta_{3}=y$.
The left and right panels correspond to cases of $c_{s}^{2}=1/3$
and $c_{s}^{2}=1$, respectively.}

\label{fig:1}
\end{figure}

As shown in figure \ref{fig:1}, for $c_{s}^{2}=1/3$, positive $y$
produces a nearly uniform enhancement of the SIGW amplitude across
the allowed $k/k_{\ast}$ range, while negative $y$ yields a global
suppression. The spectral shape remains similar to that in GR, preserving
the sharp drop beyond $k/k_{\ast}\simeq2$ due to momentum conservation.

For $c_{s}^{2}=1$, the modification is more scale-dependent: positive
$y$ not only raises the overall amplitude but also amplifies the
feature near $k/k_{\ast}\rightarrow2$, producing a pronounced peak-like
growth before the cutoff. This peak can be traced to the strong response
of the SCG kernel when both scalar modes re-enter the horizon at nearly
equal times, thereby enhancing the convolution integral in (\ref{spec}).
The appearance of such sharp-edge amplification is absent in GR, indicating
a genuine SCG-specific cubic-interaction effect.

\subsubsection{Case 2: $\beta_{3}=y$, $C_{4}^{(3)}=-2C_{5}^{(3)}=-y$}

In SCG, coefficients in $\mathcal{L}^{(2)}=c_{1}^{(2)}K_{ij}K^{ij}+c_{2}^{(2)}K^{2}+c_{3}^{(2)}R$
can generally deviate from the values in the 3+1 decomposition of
GR. We now consider the case of $\beta_{3}=y$, and for simplicity
we also assume $C_{4}^{(3)}=-2C_{5}^{(3)}=-y$. Here and in what follows,
coefficients in $\mathcal{L}^{(2)}$ that are not specified take the
same value as those in GR. For example, here we choose $c_{1}^{(2)}=-c_{2}^{(2)}=1/2$.

\begin{figure}[H]
\includegraphics[scale=0.4]{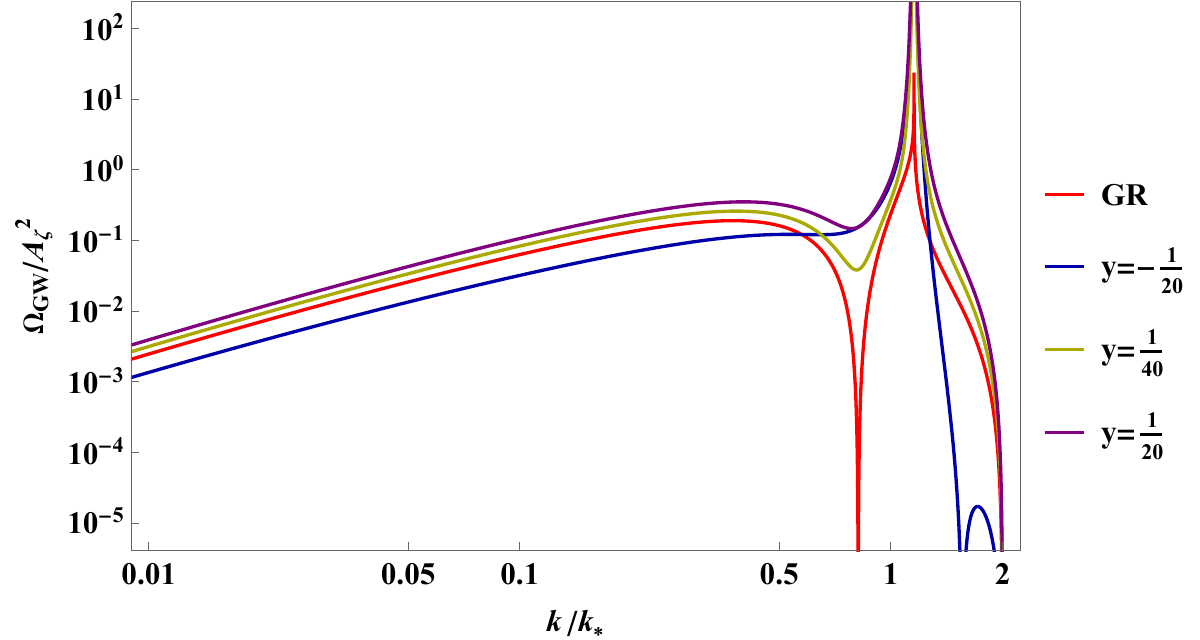}\includegraphics[scale=0.4]{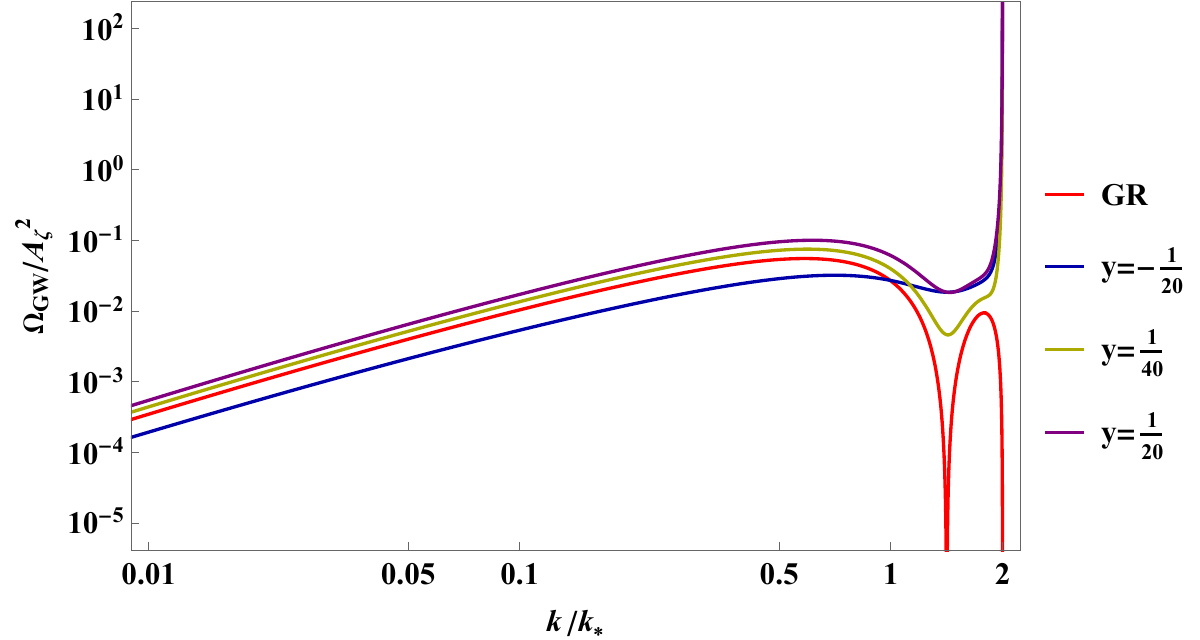}

\caption{Energy density of SIGWs with $\beta_{3}=y,C_{4}^{(3)}=-2C_{5}^{(3)}=-y$.
The left and right panels correspond to cases of $c_{s}^{2}=1/3$
and $c_{s}^{2}=1$, respectively.}

\label{fig:2}
\end{figure}

In figure \ref{fig:2}, with $c_{s}^{2}=1/3$, a positive $y$ enhances
the entire power spectrum and the shape of the spectrum is not significantly
influenced. The \textquotedblleft valley\textquotedblright{} feature
in GR around $k/k_{\ast}\sim1.4$ is smoothed out, indicating modified
interference between scalar source terms at different momenta. For
negative $y$, the suppression is strongest near $k/k_{\ast}\approx2$,
shifting the location of the minimum slightly, a sign that the SCG
Lagrangian $\mathcal{L}^{(2)}$ affects the effective phase velocities
in the kernel.

For $c_{s}^{2}=1$, the effect is more dramatic. When $k/k_{\ast}\rightarrow2$,
the spectrum tends toward a divergence. This can be traced to contributions
from terms in $I_{\mathrm{p}}$ oscillating with $\sin(ux-vx)$ at
nearly zero phase velocity difference, which in SCG are not exactly
canceled as in GR due to reduced symmetry. Physically, this signals
that, in the unitary gauge formulation of SCG, certain cubic couplings
allow \textquotedblleft slow\textquotedblright{} tensor modes to resonate
with scalar modes, enhancing the edge behavior.

\subsubsection{Case 3: $\beta_{1}=-\beta_{2}=y$, $\beta_{3}=Y$, $C_{4}^{(3)}=-2C_{5}^{(3)}=y-Y$}

\begin{figure}[H]
\includegraphics[scale=0.4]{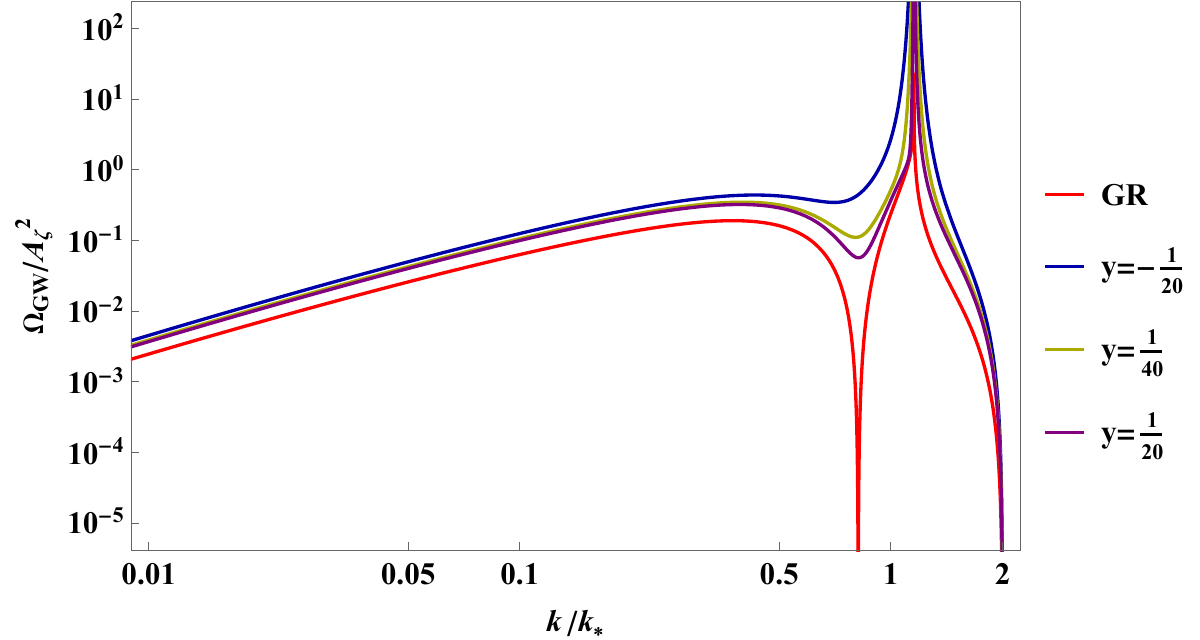}\includegraphics[scale=0.4]{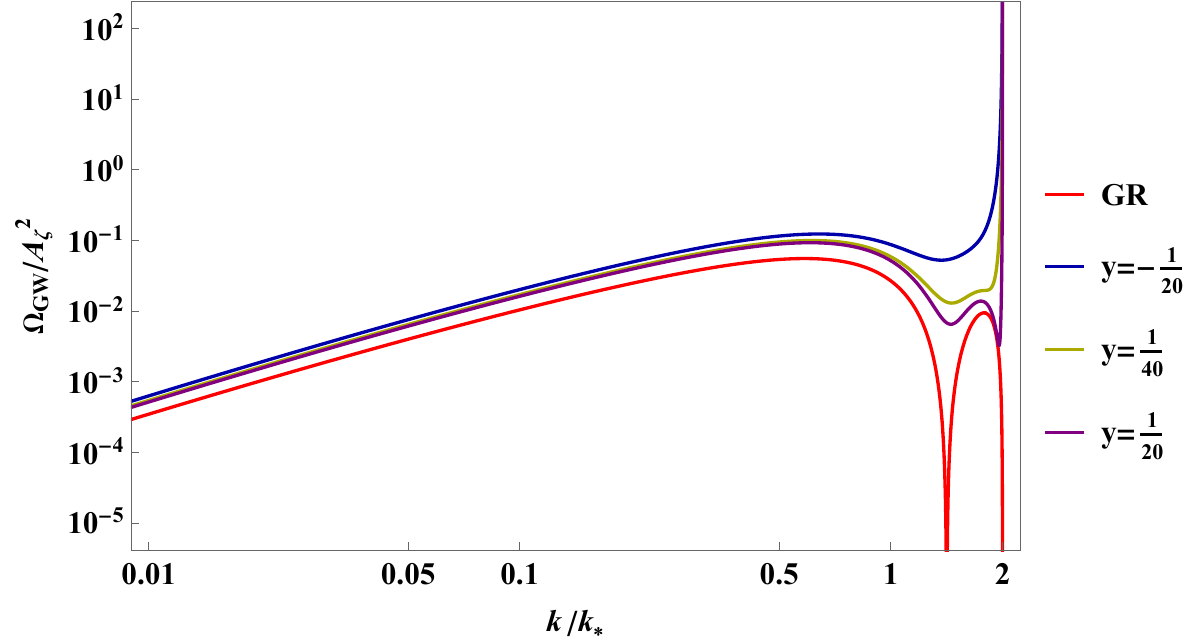}

\caption{Energy density of SIGWs with $\beta_{1}=-\beta_{2}=y$, $\beta_{3}=Y=1/18$,
$C_{4}^{(3)}=-2C_{5}^{(3)}=y-Y$. The left and right panels correspond
to cases of $c_{s}^{2}=1/3$ and $c_{s}^{2}=1$, respectively.}

\label{fig:3}
\end{figure}

We now consider the case of $\beta_{1}=-\beta_{2}=y$, $\beta_{3}=Y$,
where $y$ and $Y$ are two constants with $y\neq Y$. Figure \ref{fig:3}
illustrates the interplay between $y$ and $Y$.

When $c_{s}^{2}=1/3$, spectra retain the GR-like overall shape but
shift in amplitude depending on the sign of $y$. The interference
pattern (peaks and troughs) is slightly phase-shifted, which can be
attributed to the mismatch between the quadratic and cubic SCG coefficients.

For $c_{s}^{2}=1$, a positive $y$ again leads to edge growth near
$k/k_{\ast}\approx2$, but less violently than in Case 2, suggesting
partial cancellation between $y$- and $Y$-dependent terms in the
kernel.

\subsubsection{Case 4: $C_{3}^{(4)}=-2C_{3}^{(5)}=y$, $\frac{1}{2}C_{3}^{(1)}=-\frac{1}{3}C_{3}^{(2)}=C_{3}^{(3)}=Y=-\frac{1}{3}y$}

The last case we consider is the combination of coefficients as those
in the Horndeski theory. Unfortunately, we have not found another
choice of coefficients that leads to a power spectrum that vanishes
at large scales ($k\to0$), and most choices blow up at large scales. 

\begin{figure}[H]
\includegraphics[scale=0.4]{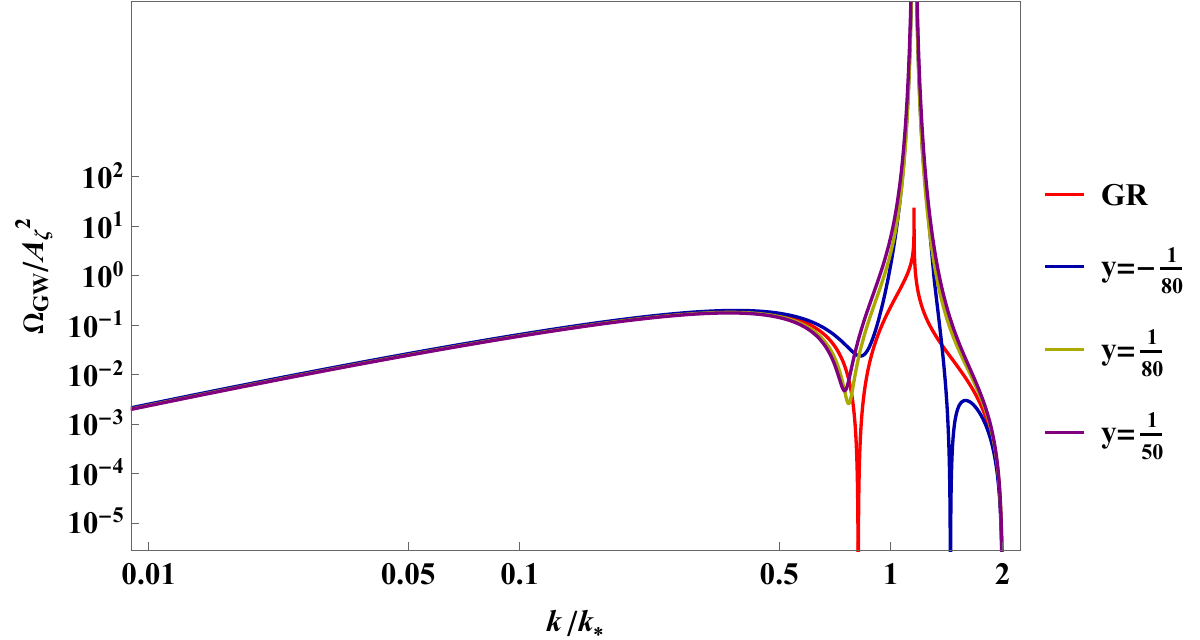}\includegraphics[scale=0.4]{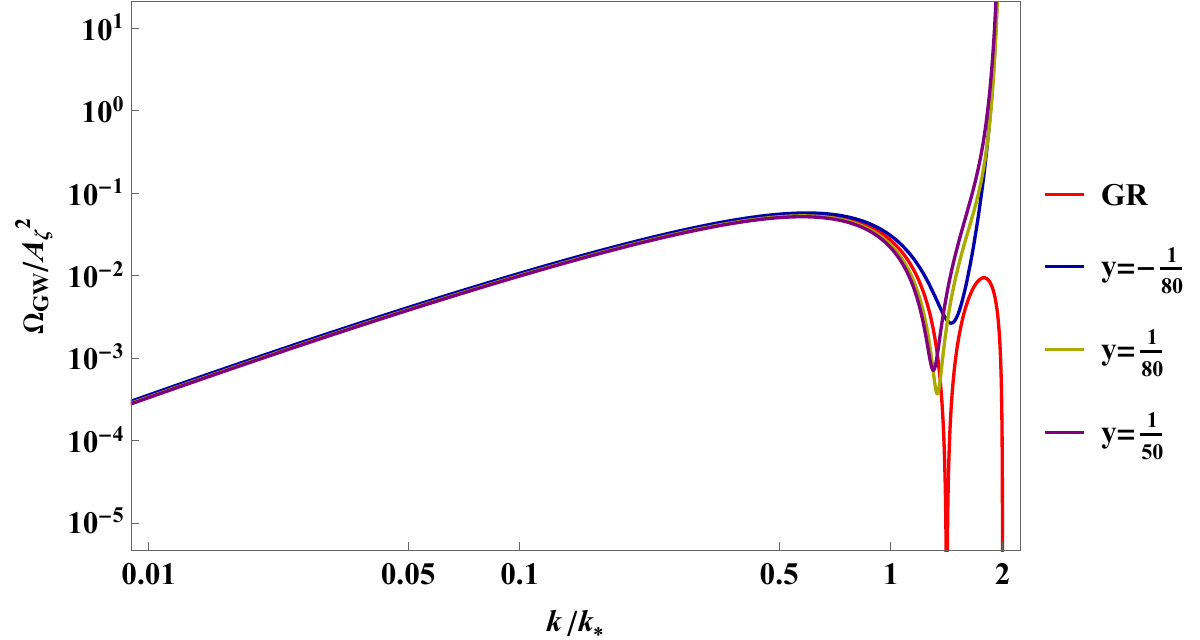}

\caption{Energy density of SIGWs with $C_{3}^{(4)}=-2C_{3}^{(5)}=y$, $\frac{1}{2}C_{3}^{(1)}=-\frac{1}{3}C_{3}^{(2)}=C_{3}^{(3)}=Y=-\frac{1}{3}y$.
The left and right panels correspond to cases of $c_{s}^{2}=1/3$
and $c_{s}^{2}=1$, respectively.}

\label{fig:4}
\end{figure}

In figure \ref{fig:4}, the Horndeski-inspired parameter choice produces
a qualitatively different outcome. For $c_{s}^{2}=1/3$, the spectrum
remains finite but exhibits a steep rise as $k/k_{\ast}\rightarrow0$,
unlike the flat IR plateau in GR. This IR growth arises because the
SCG cubic vertices in this combination leave a residual constant term
in $I_{\mathrm{o}}$ that integrates to a nonzero limit as $k\rightarrow0$.

For $c_{s}^{2}=1$, this case still avoids the sharp divergence at
$k/k_{\ast}=2$, but yields an intermediate-$k$ enhancement, reflecting
the Horndeski structure's tendency to suppress high $u,v$ interference
while boosting moderate momentum transfer.

Across all cases, SCG parameters control both the global amplitude
and the detailed interference patterns of the SIGW spectrum. The sharp-edge
enhancement near $k/k_{\ast}=2$ for $c_{s}^{2}=1$ is a robust SCG
signature whenever cubic terms fail to cancel \textquotedblleft slow\textquotedblright{}
oscillatory kernel pieces --- something absent in GR due to full
diffeomorphism invariance. For $c_{s}^{2}=1/3$, changes are smoother,
dominated by amplitude rescaling and valley suppression/shift. These
features, if observed in future high-frequency GW measurements, could
serve as direct probes of SCG operator structure.

\subsection{Log-normal primordial spectrum}

We now consider a log-normal primordial curvature spectrum 
\begin{equation}
P_{\zeta}=A_{\zeta}\frac{1}{\sqrt{2\pi}\varDelta}\exp\left(-\frac{\log^{2}(k/k_{\ast})}{2\varDelta^{2}}\right),
\end{equation}
where $\varDelta$ represents a finite-width peak. This smooth profile
removes the sharp cutoff at $k/k_{\ast}=2$, so interference patterns
from the kernel are blended over a broader range of $u,v$, and any
resonant enhancement is less singular.

Unfortunately, there are no analytical expressions for $\varOmega_{\mathrm{GR}}$
in this case. Thus we will perform the integration numerically and
give the results of 4 concrete cases of coefficients. Without loss
of generality, in the following we set $\Delta=0.1$.

\subsubsection{Case 1: $\beta_{1}=-\beta_{2}=\beta_{3}=y$}

\begin{figure}[H]
\includegraphics[scale=0.4]{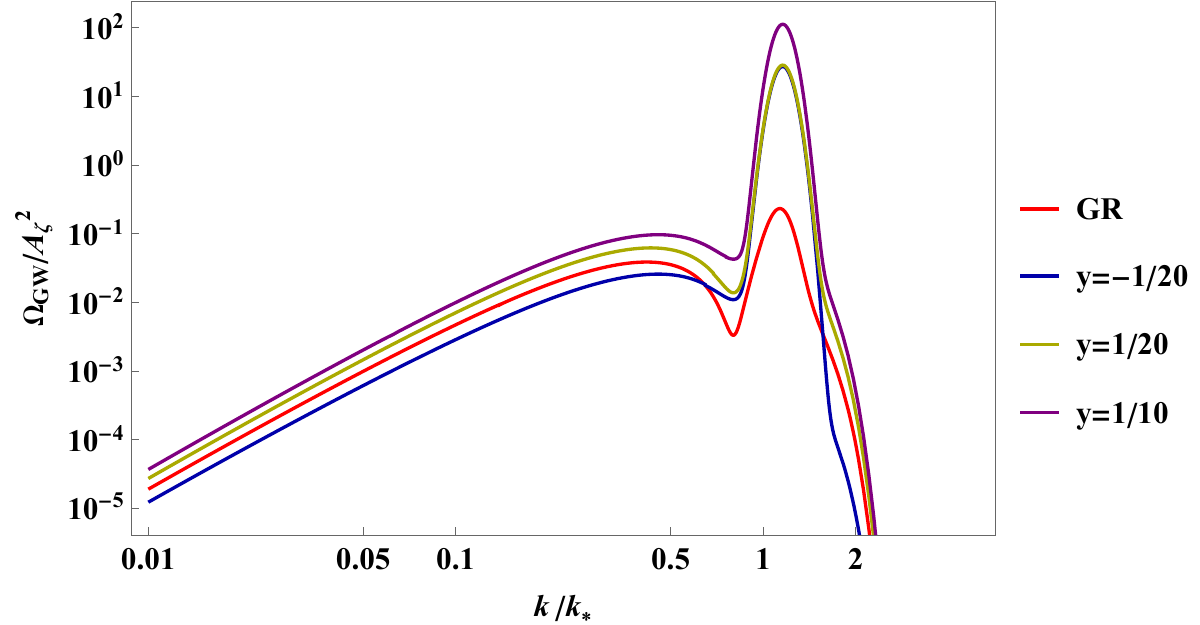}\includegraphics[scale=0.4]{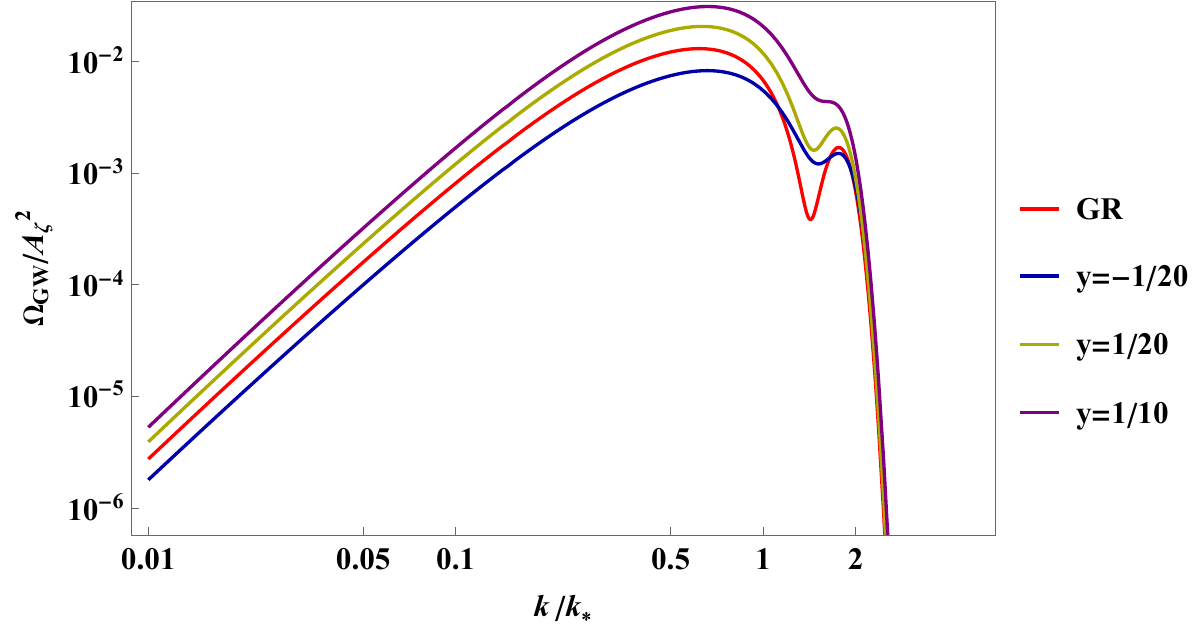}

\caption{Energy density of SIGWs with $\beta_{1}=-\beta_{2}=\beta_{3}=y$.
The left and right panels correspond to cases of $c_{s}^{2}=1/3$
and $c_{s}^{2}=1$, respectively.}

\label{fig:5}
\end{figure}

Figure \ref{fig:5} shows that for $c_{s}^{2}=1/3$, positive $y$
produces a nearly scale-independent enhancement over the GR curve,
while negative $y$ yields a mild suppression.

For $c_{s}^{2}=1$, the finite width of $\mathcal{P}_{\zeta}$ smears
the $k/k_{\ast}\approx2$ edge peak into a broader shoulder, reducing
the visual sharpness but still yielding a noticeable high-$k$ excess
for positive $y$.

\subsubsection{Case 2: $\beta_{3}=y$, $C_{4}^{(3)}=-2C_{5}^{(3)}=-y$}

In contrast to Case 1, we find that the spectrum diverges when $c_{s}^{2}=1/3$.
Therefore in the following cases we focus on the case of $c_{s}^{2}=1$.

\begin{figure}[H]
\begin{centering}
\includegraphics[scale=0.4]{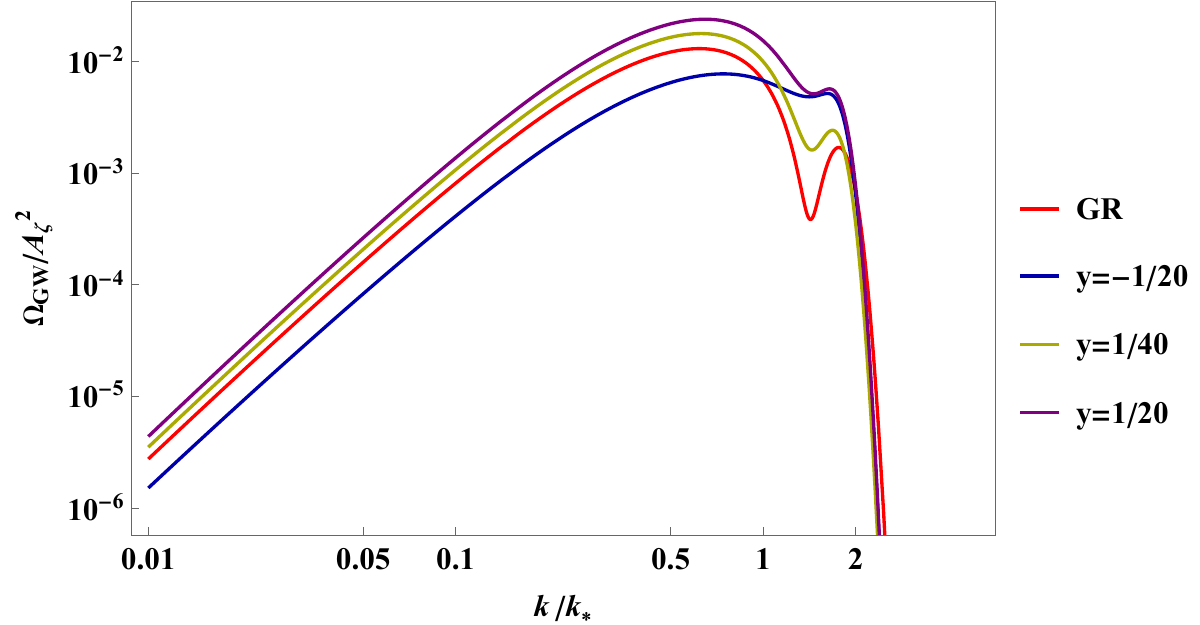}
\par\end{centering}
\caption{Energy density of SIGWs with $\beta_{3}=y$, $C_{4}^{(3)}=-2C_{5}^{(3)}=-y$.
We have chosen $c_{s}^{2}=1$.}

\label{fig:6}
\end{figure}

In figure \ref{fig:6}, positive $y$ lifts the spectrum primarily
in the central $k$-range $0.5\lesssim k/k_{\ast}\lesssim1.5$, leaving
the tails closer to the GR baseline. This suggests that the SCG modifications
here act most strongly when both scalar source modes have wavenumbers
near the peak of $\mathcal{P}_{\zeta}$, consistent with kernel terms
in $I_{\mathrm{o}}$ that weight near-equal $u,v$ most heavily.

\subsubsection{Case 3: $\beta_{1}=-\beta_{2}=y$, $\beta_{3}=Y$, $C_{4}^{(3)}=-2C_{5}^{(3)}=y-Y$}

\begin{figure}[H]
\begin{centering}
\includegraphics[scale=0.4]{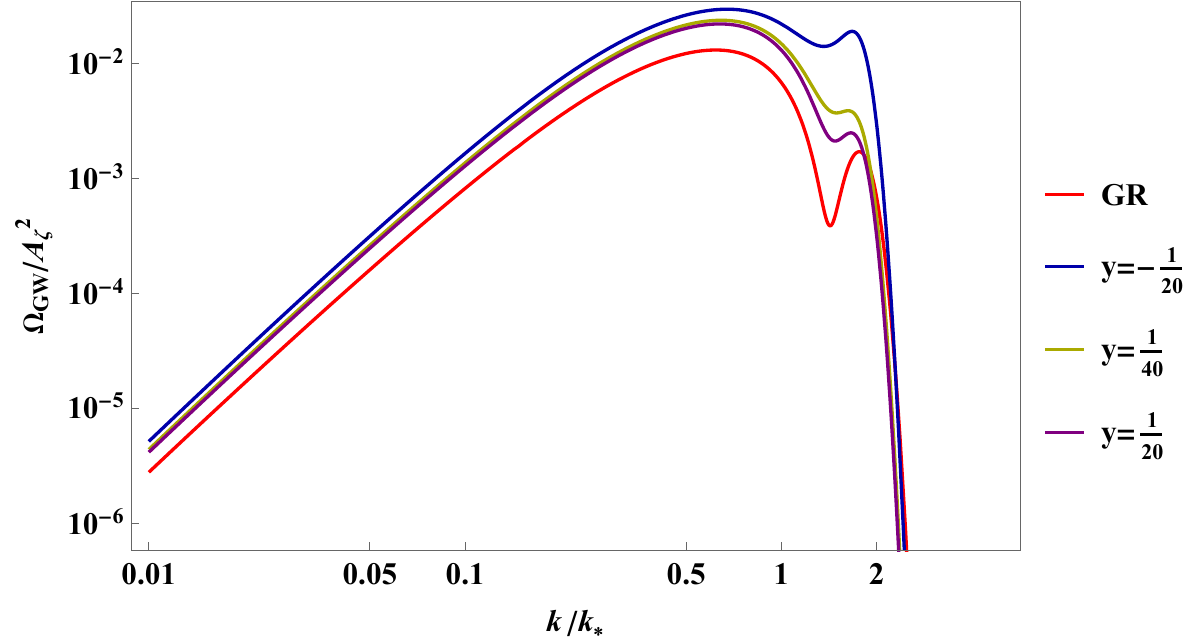}
\par\end{centering}
\caption{Energy density of SIGWs with $\beta_{1}=-\beta_{2}=y$, $\beta_{3}=Y=1/18$,
$C_{4}^{(3)}=-2C_{5}^{(3)}=y-Y$. We have chosen $c_{s}^{2}=1$.}

\label{fig:7}
\end{figure}

As shown in figure \ref{fig:7}, adjusting $y$ shifts the amplitude
up or down, while $Y$ modulates the slope on the high-$k$ side.
This asymmetry relative to the peak position reflects the interplay
between quadratic and cubic modifications in redistributing power
between configurations with $u\approx v$ and those with larger momentum
asymmetry.

\subsubsection{Case 4: $C_{3}^{(4)}=-2C_{3}^{(5)}=y$, $\frac{1}{2}C_{3}^{(1)}=-\frac{1}{3}C_{3}^{(2)}=C_{3}^{(3)}=Y=-\frac{1}{3}y$}

\begin{figure}[H]
\begin{centering}
\includegraphics[scale=0.4]{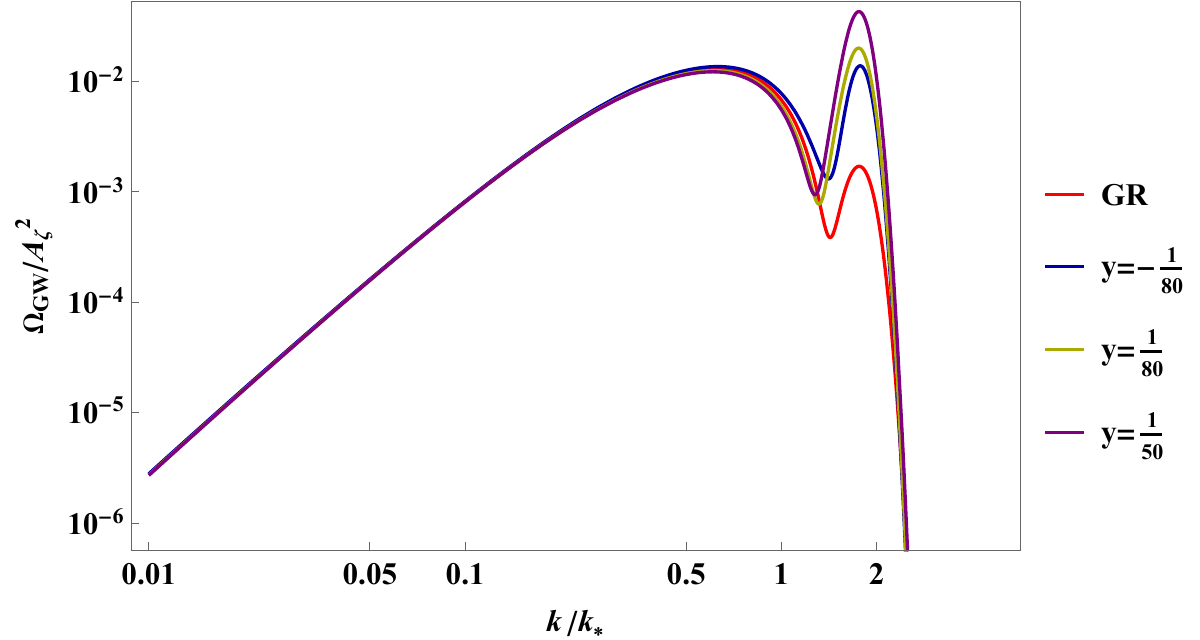}
\par\end{centering}
\caption{Energy density of SIGWs with $C_{3}^{(4)}=-2C_{3}^{(5)}=y$, $\frac{1}{2}C_{3}^{(1)}=-\frac{1}{3}C_{3}^{(2)}=C_{3}^{(3)}=Y=-\frac{1}{3}y$.
We have chosen $c_{s}^{2}=1$.}

\label{fig:8}
\end{figure}

In figure \ref{fig:8}, this combination yields a spectrum that is
more symmetric about $k/k_{\ast}=1$ than the previous cases, with
moderate enhancement at both low and high $k$ relative to the peak.
The Horndeski-like structure appears to \textquotedblleft flatten\textquotedblright{}
the spectrum, suppressing the pronounced shoulders or valleys seen
in other parameter choices.

Compared with the monochromatic spectrum case, the log-normal input
smooths sharp features and reduces  sensitivity to the exact cutoff
at $k/k_{\ast}=2$. SCG effects manifest mainly as amplitude shifts
and mild shape distortions, but high-$k$ shoulders can still appear
when cubic couplings preferentially amplify near-equal-$u,v$ configurations.
Taking into account the observations, the log-normal case suggests
that distinguishing SCG from GR would require precise measurement
of the shape of the SIGW bump, not just its amplitude.

\section{Conclusion \label{sec:con}}

We have carried out a systematic computation of scalar-induced gravitational
waves in spatially covariant gravity, focusing on polynomial-type
Lagrangians up to $d=3$ with $d$ the total number of derivatives
in each monomial. 

In Sec. \ref{sec:linear}, we studied the background evolution and
linear perturbations of the scalar and tensor modes in a cosmological
background. Due to the breaking of time reparametrization symmetry
of SCG, we are only allowed to work in unitary gauge instead of the
Newtonian gauge usually adopted in calculating SIGWs in GR. This choice
not only affects the formal structure of the perturbation equations
but also introduces new couplings that play a crucial role in shaping
the properties of SIGWs.

In Sec. \ref{sec:cubeom}, we derived the cubic action involving one
tensor and two scalar modes and obtained the general equations of
motion for SIGWs. From this, we extracted the source term and kernel
function, whose time dependence encodes the deviations of SCG from
GR. By considering power-law time evolution of the coefficients, we
presented explicit results and restricted attention to operators yielding
luminal tensor propagation. This restriction ensured consistency with
observational bounds and avoided the spurious late-time divergences
that can otherwise appear in $\varOmega_{\mathrm{GR}}$. Compared
with GR, the SCG kernels contain additional oscillatory structures
and modified amplitude scaling, providing characteristic imprints
in the resulting spectra.

In Sec. \ref{sec:spec}, we evaluated the fractional energy density
of SIGWs in two representative cases of the primordial curvature perturbation
spectrum. One is the monochromatic spectrum, the other is the log-normal
spectrum. We show the energy density $\varOmega_{\mathrm{GW}}$ for
a couple of choices of coefficients. Our results show that SCG can
produce significant departures from GR predictions, including scale-dependent
amplitude shifts, enhanced edge behavior near $k/k_{\ast}\approx2$,
and modified spectral shapes. These features originate from SCG cubic
interactions and the modified kernel function, which affect both the
generation and propagation of SIGWs and remain even when deviations
from GR are small. 

The results presented in this work highlight the potential of stochastic
GW background measurements to serve as sensitive probes of Lorentz-violating
modifications to gravity. Future observations, ranging from pulsar
timing arrays to space-based interferometers, may thus be capable
of detecting or constraining these deviations. Extending the present
work to include parity-violating operators, higher-order derivative
terms, or couplings to additional fields would further broaden the
phenomenology and provide valuable directions for testing gravity
with GWs.
\begin{acknowledgments}
We would like to thank Jia-Xi Feng and Fengge Zhang for valuable discussions.
X.G. is supported by the National Natural Science Foundation of China
(NSFC) under Grants No. 12475068 and No. 11975020 and the Guangdong
Basic and Applied Basic Research Foundation under Grant No. 2025A1515012977.
\end{acknowledgments}

\appendix

\section{Decomposition of the tensor perturbation \label{app:decten}}

The tensor perturbations can be decomposed into the polarization modes
as
\begin{equation}
\gamma_{ij}(\bm{k},t)=\sum_{\lambda}e_{ij}^{\lambda}(\bm{k})\gamma_{\lambda}(\bm{k},t),
\end{equation}
where $\lambda=+,\times$ denotes the two polarization modes.

The polarization tensors are defined in terms of the polarization
vectors as 
\begin{align}
e_{ij}^{+}(\bm{k}) & =\frac{1}{\sqrt{2}}[e_{i}(\bm{k})e_{j}(\mathbf{k})-\bar{e}_{i}(\bm{k})\bar{e}_{j}(\bm{k})],\\
e_{ij}^{\times}(\bm{k}) & =\frac{1}{\sqrt{2}}[e_{i}(\bm{k})e_{j}(\bm{k})+\bar{e}_{i}(\bm{k})\bar{e}_{j}(\bm{k})],
\end{align}
where $e_{i}$ satisfies $e_{i}(\bm{k})e^{i}(\bm{k})=1$ and $e_{i}(\bm{k})\bar{e}^{i}(\bm{k})=0$.
The polarization tensors satisfy
\begin{equation}
e_{ij}(\bm{k})e^{+ij}(-\bm{k})=1,\quad e_{ij}^{+}(\bm{k})e^{\times ij}(-\bm{k})=0,
\end{equation}
and
\begin{align}
\text{symmetric:\ \ } & e_{ij}^{\lambda}(\bm{k})=e_{ji}^{\lambda}(\bm{k}),\\
\text{transverse:\ \ } & e_{ij}^{\lambda}(\bm{k})k^{i}=0,\\
\text{traceless:\ \ } & e_{ij}^{\lambda}(\bm{k})\delta^{ij}=0,\\
\text{real:\ \ } & e_{ij}^{\lambda}(\bm{k})=e_{ij}^{\lambda}(-\bm{k}).
\end{align}

Note we have
\begin{align}
\gamma^{ij}(\bm{k},t)\gamma_{ij}(-\bm{k},t) & =\sum_{\lambda,\rho=+,\times}e_{ij}^{\lambda}(\bm{k})\gamma_{\lambda}(\bm{k},t)e_{ij}^{\rho}(-\bm{k})\gamma_{\rho}(-\bm{k},t)\nonumber \\
 & =\sum_{\lambda=+,\times}\gamma_{\lambda}(\bm{k},t)\gamma_{\lambda}(-\bm{k},t).
\end{align}

\section{Coefficients of quadratic actions \label{app:coeffsca}}

The explicit expressions of the coefficients in \ref{S2zAB} are given
by
\begin{align}
\mathcal{C}_{\zeta'\zeta'} & =\frac{3J}{2\mathcal{H}^{2}},\\
\mathcal{C}_{\zeta\zeta'} & =\frac{1}{\mathcal{H}}\left[K+\frac{9}{2}J+\frac{k^{2}}{\mathcal{H}^{2}}\left(4(\frac{\mathcal{H}}{a})^{3}c_{4}^{(3)}+12(\frac{\mathcal{H}}{a})^{3}c_{5}^{(3)}\right)\right],\\
\mathcal{C}_{\zeta\zeta} & =-\frac{9}{2}c'{}_{1,0}^{(0)}+\frac{3}{2}K+\frac{27}{4}J+\frac{k^{2}}{\mathcal{H}^{2}}\left(2(\frac{\mathcal{H}}{a})^{2}c_{3}^{(2)}+2(\frac{\mathcal{H}}{a})^{3}c_{4}^{(3)}+6(\frac{\mathcal{H}}{a})^{3}c_{5}^{(3)}\right),\\
\mathcal{C}_{A\zeta'} & =-\frac{3J}{\mathcal{H}},\\
\mathcal{C}_{A\zeta} & =\frac{k^{2}}{\mathcal{H}^{2}}\left(4(\frac{\mathcal{H}}{a})^{2}c_{3}^{(2)}\right),\\
\mathcal{C}_{AA} & =\frac{U}{2}+\frac{3}{2}J,\\
\mathcal{C}_{AB} & =-\mathcal{H}\left(\frac{k^{2}}{\mathcal{H}^{2}}J\right),\\
\mathcal{C}_{B\zeta'} & =\frac{k^{2}}{\mathcal{H}^{2}}J,\\
\mathcal{C}_{B\zeta} & =\mathcal{H}\left[\frac{k^{4}}{\mathcal{H}^{4}}\left(2(\frac{\mathcal{H}}{a})^{3}c_{4}^{(3)}+4(\frac{\mathcal{H}}{a})^{3}c_{5}^{(3)}\right)\right],\\
\mathcal{C}_{BB} & =\mathcal{H}^{2}\left[\frac{k^{4}}{\mathcal{H}^{4}}\left(\frac{J}{6}+\frac{I}{6}\right)\right],
\end{align}
where 
\begin{align}
K & =9(\frac{\mathcal{H}}{a})c_{1}^{(1)}+9(\frac{\mathcal{H}}{a})^{2}c_{1}^{(2)}+27(\frac{\mathcal{H}}{a})^{2}c_{2}^{(2)},\\
I & =4(\frac{\mathcal{H}}{a})^{2}c_{1}^{(2)}+12(\frac{\mathcal{H}}{a})^{3}c_{1}^{(3)}+12(\frac{\mathcal{H}}{a})^{3}c_{2}^{(3)},\\
J & =2(\frac{\mathcal{H}}{a})^{2}c_{1}^{(2)}+6(\frac{\mathcal{H}}{a})^{2}c_{2}^{(2)}+6(\frac{\mathcal{H}}{a})^{3}c_{1}^{(3)}+18(\frac{\mathcal{H}}{a})^{3}c_{2}^{(3)}+54(\frac{\mathcal{H}}{a})^{3}c_{3}^{(3)},\\
U & =2c_{1,0}^{'(0)}+c_{1,0}^{''(0)}.
\end{align}

Solution to $A,B$ are
\begin{align}
A_{k} & =\frac{3IJ}{\mathcal{H}\mathcal{J}}\zeta_{\mathbf{k}}'-\frac{k^{2}}{\mathcal{H}^{2}}\frac{4(\frac{\mathcal{H}}{a})^{2}c_{2}^{(2)}(I+J)+6\left((\frac{\mathcal{H}}{a})^{3}c_{4}^{(3)}+2(\frac{\mathcal{H}}{a})^{3}c_{5}^{(3)}\right)J}{\mathcal{J}}\zeta_{\mathbf{k}}\\
B_{k} & =-\frac{1}{k^{2}}\frac{3JU}{\mathcal{J}}\zeta_{\mathbf{k}}'-\frac{6\left((\frac{\mathcal{H}}{a})^{3}c_{4}^{(3)}+2(\frac{\mathcal{H}}{a})^{3}c_{5}^{(3)}\right)(3J+U)+12J(\frac{\mathcal{H}}{a})^{2}c_{3}^{(2)}}{\mathcal{H}\mathcal{J}}\zeta_{\mathbf{k}}
\end{align}
where 
\begin{equation}
\mathcal{J}=3IJ+U(I+J)
\end{equation}

Coefficients in (\ref{S2zeta}) are
\begin{eqnarray}
\mathcal{D_{\zeta\zeta}} & = & \frac{3}{4}(2K+9J-6c_{1,0}^{'(0)})+2\frac{k^{2}}{\mathcal{H}^{2}}\left((\frac{\mathcal{H}}{a})^{2}c_{3}^{(2)}+(\frac{\mathcal{H}}{a})^{3}c_{4}^{(3)}+3(\frac{\mathcal{H}}{a})^{3}c_{5}^{(3)}\right)\nonumber \\
 &  & -\frac{k^{4}}{\mathcal{H}^{4}\mathcal{J}}\Bigg[24(\frac{\mathcal{H}}{a})^{2}c_{3}^{(2)}\left((\frac{\mathcal{H}}{a})^{3}c_{4}^{(3)}+2(\frac{\mathcal{H}}{a})^{3}c_{5}^{(3)}\right)+8\left((\frac{\mathcal{H}}{a})^{2}c_{3}^{(2)}\right)^{2}(I+J)\nonumber \\
 &  & \qquad+6\left((\frac{\mathcal{H}}{a})^{3}c_{4}^{(3)}+2(\frac{\mathcal{H}}{a})^{3}c_{5}^{(3)}\right)^{2}(3J+U)\Bigg],
\end{eqnarray}
\begin{eqnarray}
D_{\zeta\zeta'} & = & \frac{1}{\mathcal{H}}\left(K+\frac{9}{2}J\right)\nonumber \\
 &  & +\frac{k^{2}}{\mathcal{H}^{3}}\left[4\left((\frac{\mathcal{H}}{a})^{3}c_{4}^{(3)}+3(\frac{\mathcal{H}}{a})^{3}c_{5}^{(3)}\right)+\frac{12IJ}{\mathcal{J}}(\frac{\mathcal{H}}{a})^{2}c_{3}^{(2)}-\frac{6UJ}{\mathcal{J}}\left((\frac{\mathcal{H}}{a})^{3}c_{4}^{(3)}+2(\frac{\mathcal{H}}{a})^{3}c_{5}^{(3)}\right)\right],
\end{eqnarray}
\begin{equation}
D_{\zeta'\zeta'}=\frac{1}{\mathcal{H}^{2}}\frac{3IJU}{2\mathcal{J}}.
\end{equation}

\section{Coefficients of Cubic action \label{app:coeffS3}}

Coefficients in (\ref{S3}) are
\begin{eqnarray}
\mathcal{F}_{1} & = & \frac{J}{\mathcal{H}},\\
\mathcal{F}_{2} & = & -\frac{l^{2}}{\mathcal{H}^{2}}\left(2(\frac{\mathcal{H}}{a})^{2}c_{1}^{(2)}+2(\frac{\mathcal{H}}{a})^{2}c_{2}^{(2)}+6(\frac{\mathcal{H}}{a})^{3}c_{1}^{(3)}+10(\frac{\mathcal{H}}{a})^{3}c_{2}^{(3)}+18(\frac{\mathcal{H}}{a})^{3}c_{3}^{(3)}\right)-\frac{I(\mathbf{k}\cdot\mathbf{l})}{4\mathcal{H}^{2}},\\
\mathcal{F}_{3} & = & -\frac{4}{\mathcal{H}^{2}}\left((\frac{\mathcal{H}}{a})^{2}c_{3}^{(2)}\right),\\
\mathcal{F}_{4} & = & -\frac{l^{2}}{\mathcal{H}^{3}}\left(\frac{5}{2}(\frac{\mathcal{H}}{a})^{3}c_{4}^{(3)}+4(\frac{\mathcal{H}}{a})^{3}c_{5}^{(3)}\right),\\
\mathcal{F}_{5} & = & -\frac{l^{2}}{\mathcal{H}^{3}}\left(\frac{3}{2}(\frac{\mathcal{H}}{a})^{3}c_{4}^{(3)}+4(\frac{\mathcal{H}}{a})^{3}c_{5}^{(3)}\right)-\frac{(\mathbf{k}\cdot\mathbf{l})}{\mathcal{H}^{3}}\left((\frac{\mathcal{H}}{a})^{3}c_{4}^{(3)}\right)+\frac{k^{2}}{\mathcal{H}^{3}}\left(\frac{1}{2}(\frac{\mathcal{H}}{a})^{3}c_{4}^{(3)}\right),\\
\mathcal{F}_{6} & = & -\frac{J}{\mathcal{H}^{2}},\\
\mathcal{F}_{7} & = & -\frac{1}{\mathcal{H}^{2}}\left(2(\frac{\mathcal{H}}{a})^{2}c_{3}^{(2)}+2(\frac{\mathcal{H}}{a})^{3}c_{4}^{(3)}+6(\frac{\mathcal{H}}{a})^{3}c_{5}^{(3)}\right),\\
\mathcal{F}_{8} & = & -\frac{1}{\mathcal{H}^{3}}\left(4(\frac{\mathcal{H}}{a})^{3}c_{4}^{(3)}+12(\frac{\mathcal{H}}{a})^{3}c_{5}^{(3)}\right),
\end{eqnarray}
and
\begin{eqnarray}
\mathcal{G}_{1} & = & -\frac{1}{\mathcal{H}^{2}}\left((\frac{\mathcal{H}}{a})^{2}c_{1}^{(2)}+6(\frac{\mathcal{H}}{a})^{3}c_{1}^{(3)}+6(\frac{\mathcal{H}}{a})^{3}c_{2}^{(3)}\right),\\
\mathcal{G}_{2} & = & -\frac{3}{2}\frac{(\mathbf{k}\cdot\mathbf{l})}{\mathcal{H}^{3}}(\frac{\mathcal{H}}{a})^{3}c_{1}^{(3)}+\frac{l^{2}}{\mathcal{H}^{3}}\left(\frac{3}{2}(\frac{\mathcal{H}}{a})^{3}c_{1}^{(3)}+(\frac{\mathcal{H}}{a})^{3}c_{2}^{(3)}\right),\\
\mathcal{G}_{3} & = & \frac{3}{\mathcal{H}^{3}}\left((\frac{\mathcal{H}}{a})^{3}c_{1}^{(3)}+(\frac{\mathcal{H}}{a})^{3}c_{2}^{(3)}\right),\\
\mathcal{G}_{4} & = & \frac{3I}{4\mathcal{H}^{2}},\\
\mathcal{G}_{5} & = & \frac{1}{\mathcal{H}^{3}}\left((\frac{\mathcal{H}}{a})^{3}c_{4}^{(3)}\right).
\end{eqnarray}

Coefficients in (\ref{eq:4}) are 
\begin{eqnarray}
a^{2}\mathcal{T}_{\zeta\zeta} & = & \frac{1}{2a^{2}\mathcal{J}^{2}}3\mathcal{H}^{2}(3J+U)(c_{4}^{(3)}+2c_{5}^{(3)}){}^{2}\left(4JU\left(|\mathbf{k}-\mathbf{q}|^{2}+q^{2}\right)-i(3J+U)\left(3k^{2}-4\left(|\mathbf{k}-\mathbf{q}|^{2}+q^{2}\right)\right)\right)\nonumber \\
 &  & +\frac{1}{\mathcal{J}^{2}}2c_{3}^{(2)}{}^{2}\Big[4J^{2}U\left(|\mathbf{k}-\mathbf{q}|^{2}+q^{2}\right)+IJ\left(-9Jk^{2}+12J\left(|\mathbf{k}-\mathbf{q}|^{2}+q^{2}\right)+8U\left(|\mathbf{k}-\mathbf{q}|^{2}+q^{2}\right)\right)\nonumber \\
 &  & \hspace*{4em}-4(3J+U)\left(|\mathbf{k}-\mathbf{q}|^{2}+q^{2}\right)\Big]\nonumber \\
 &  & -\frac{1}{a\mathcal{J}^{2}}6HJc_{3}^{(2)}(c_{4}^{(3)}+2c_{5}^{(3)})\left(-4JU\left(|\mathbf{k}-\mathbf{q}|^{2}+q^{2}\right)+I(3J+U)\left(3k^{2}-4\left(|\mathbf{k}-\mathbf{q}|^{2}+q^{2}\right)\right)\right)\nonumber \\
 &  & -2a^{2}c_{3}^{(2)}-2a\mathcal{H}(c_{4}^{(3)}+3c_{5}^{(3)}),
\end{eqnarray}
\begin{align}
a^{2}\mathcal{T}_{\zeta'\zeta} & =\frac{a}{4\mathcal{H}q^{2}\mathcal{J}^{2}}\times\bigg\{\left(6ac_{3}^{(2)}IJ\left(JU\left(3k^{2}+4q^{2}\right)+4q^{2}I(3J+U)\right)\right)\nonumber \\
 & +\mathcal{H}\bigg(c_{4}^{(3)}\left(-4J^{2}q^{2}U^{2}+IJU(3J+U)\left(9k^{2}+4q^{2}\right)-8q^{2}(3J+U)^{2}\right)\nonumber \\
 & +6c_{5}^{(3)}I(3J+U)\left(JU\left(3k^{2}+4q^{2}\right)+4q^{2}I(3J+U)\right)\bigg)\bigg\},
\end{align}
\begin{equation}
\mathcal{T}_{\zeta\zeta'}=T_{\zeta\zeta'}(q\leftrightarrow|\mathbf{k}-\mathbf{q}|),
\end{equation}
\begin{equation}
a^{2}\mathcal{T}_{\zeta'\zeta'}=-\frac{9a^{4}k^{2}IJ^{2}U^{2}}{8\mathcal{H}^{2}|\mathbf{k}-\mathbf{q}|^{2}q^{2}\mathcal{J}^{2}}.
\end{equation}

Coefficients of terms with $\gamma'$ are
\begin{eqnarray}
a^{2}\mathcal{X}_{\zeta\zeta} & = & -\frac{1}{2a^{5}\mathcal{H}\mathcal{J}^{2}}\times\bigg\{72a^{2}\mathcal{H}^{3}Jc_{3}^{(2)}{}^{2}\left(Jc_{1}^{(3)}\left(3k^{2}-l^{2}-q^{2}\right)+(2c_{1}^{(3)}+2c_{2}^{(3)})\left(l^{2}+q^{2}\right)I\right)\nonumber \\
 &  & \quad+12a^{5}IJ(J+I)c_{3}^{(2)}{}^{2}\left(l^{2}+q^{2}\right)+9a^{3}\mathcal{H}^{2}IJ(3J+U)\left(l^{2}+q^{2}\right)(c_{4}^{(3)}+2c_{5}^{(3)}){}^{2}\nonumber \\
 &  & \quad+a^{6}\mathcal{H}(JU+I(3J+U))\left(-2c_{4}^{(3)}JU+(7c_{4}^{(3)}+18c_{5}^{(3)})(3J+U)I\right)\nonumber \\
 &  & \quad+6a^{4}c_{2}^{(3)}\mathcal{H}I\left(c_{4}^{(3)}+2c_{5}^{(3)}\right)(J(6J+U)+I(3J+U))\left(l^{2}+q^{2}\right)+18a^{7}c_{3}^{(2)}IJ(JU+I(3J+U))\nonumber \\
 &  & \quad+18\mathcal{H}^{5}(3J+U)(c_{4}^{(3)}+2c_{5}^{(3)}){}^{2}\left(3c_{1}^{(3)}\left(J\left(3k^{2}-l^{2}-q^{2}\right)+U\left(k^{2}-l^{2}-q^{2}\right)\right)-2c_{2}^{(3)}U\left(l^{2}+q^{2}\right)\right)\nonumber \\
 &  & \quad+72a\mathcal{H}^{4}c_{2}^{(3)}(c_{4}^{(3)}+2c_{5}^{(3)})\Big[c_{1}^{(3)}\Big(J\left(3I\left(l^{2}+q^{2}\right)+3k^{2}U-2U\left(l^{2}+q^{2}\right)\right)+IU\left(l^{2}+q^{2}\right)\nonumber \\
 &  & \hspace*{4em}+J^{2}\left(9k^{2}-3\left(l^{2}+q^{2}\right)\right)\Big)+c_{2}^{(3)}\left(l^{2}+q^{2}\right)(I(3J+U)-JU)\Big]\bigg\},
\end{eqnarray}
\begin{eqnarray}
a^{2}\mathcal{X}_{\zeta'\zeta} & = & -\frac{3}{8a^{2}\mathcal{H}^{2}q^{2}\mathcal{J}^{2}}\bigg\{24a\mathcal{H}^{3}c_{2}^{(3)}UJ\left(2Ic_{1}^{(3)}\left(l^{2}+q^{2}\right)+2Ic_{2}^{(3)}\left(l^{2}+q^{2}\right)+Jc_{1}^{(3)}\left(3k^{2}-l^{2}-q^{2}\right)\right)\nonumber \\
 &  & \quad+3a^{6}IJU\left(I(3J+U)+JU\right)-6a^{3}\mathcal{H}IJ(c_{4}^{(3)}+2c_{5}^{(3)})\left(Iq^{2}(3J+U)-Jl^{2}U\right)\nonumber \\
 &  & \quad+4a^{4}IJc_{2}^{(3)}\left(-3IJq^{2}+Il^{2}U+Jl^{2}U\right)\nonumber \\
 &  & \quad+12\mathcal{H}^{4}U\left(c_{4}^{(3)}+2c_{5}^{(3)}\right)\Big[Jc_{1}^{(3)}\left(6Iq^{2}+U\left(3k^{2}-3l^{2}-q^{2}\right)\right)+2Iq^{2}c_{2}^{(3)}(3J+U)\nonumber \\
 &  & \hspace*{6em}+2Iq^{2}c_{1}^{(3)}U+J^{2}c_{1}^{(3)}\left(9k^{2}-3\left(l^{2}+q^{2}\right)\right)-2Jl^{2}c_{2}^{(3)}U\Big]\bigg\},
\end{eqnarray}
\begin{equation}
a^{2}\mathcal{X}_{\zeta\zeta'}=\mathcal{X}_{\zeta'\zeta}(q\leftrightarrow l),
\end{equation}
\begin{eqnarray}
a^{2}\mathcal{X}_{\zeta'\zeta'} & = & \frac{9aJU}{8\mathcal{H}^{3}l^{2}q^{2}\mathcal{J}^{2}}\Big\{ a^{3}I^{2}J\left(l^{2}+q^{2}\right)\nonumber \\
 &  & \qquad-2\mathcal{H}^{3}U\left[2I\left(c_{1}^{(3)}+c_{2}^{(3)}\right)\left(l^{2}+q^{2}\right)+Jc_{1}^{(3)}\left(3k^{2}-l^{2}-q^{2}\right)\right]\Big\}
\end{eqnarray}
where $l^{2}=\left|\bm{k}-\bm{q}\right|^{2}$.

\section{Source function $f$ \label{app:srcfun}}

The explicit expressions of $m$'s in (\ref{eq:f}) are
\begin{eqnarray}
m_{1}^{(0)} & = & -\frac{1}{2\tilde{\mathcal{J}}}\Big[10C_{4}^{(3)}\tilde{J}\tilde{U}+12C_{5}^{(3)}\tilde{J}\tilde{U}-17C_{4}^{(3)}\tilde{I}(3\tilde{J}+\tilde{U})\nonumber \\
 &  & \qquad-42C_{5}^{(3)}\tilde{I}(3\tilde{J}+\tilde{U})-C_{3}^{(2)}(42\tilde{I}\tilde{J}-4\tilde{I}\tilde{U}-4\tilde{J}\tilde{U})\Big],
\end{eqnarray}
\begin{eqnarray}
m_{1}^{(2)} & = & \frac{k^{2}}{2\tilde{\mathcal{J}}^{2}}\Big\{3(C_{4}^{(3)}+2C_{5}^{(3)})^{2}(3\tilde{J}+\tilde{U})\Big[-4(15C_{2}^{(3)}-\tilde{J})\tilde{U}(u^{2}+v^{2})\nonumber \\
 &  & \hspace*{4em}+9\tilde{I}\tilde{J}(-1+3u^{2}+3v^{2})+\tilde{I}\tilde{U}(-3+4u^{2}+4v^{2})\Big]\nonumber \\
 &  & \quad+90C_{1}^{(3)}\bigl(2C_{3}^{(2)}\tilde{J}+(C_{4}^{(3)}+2C_{5}^{(3)})(3\tilde{J}+\tilde{U})\bigr)\Big[-2C_{3}^{(2)}\tilde{J}(-3+u^{2}+v^{2})+4C_{3}^{(2)}\tilde{I}(u^{2}+v^{2})\nonumber \\
 &  & \hspace*{4em}-3(C_{4}^{(3)}+2C_{5}^{(3)})\bigl(\tilde{J}(-3+u^{2}+v^{2})+\tilde{U}(-1+u^{2}+v^{2})\bigr)\Big]\nonumber \\
 &  & \quad+4(C_{3}^{(2)})^{2}\Big[4\tilde{J}^{2}\tilde{U}(u^{2}+v^{2})+\tilde{I}^{2}(27\tilde{J}+4\tilde{U})(u^{2}+v^{2})\nonumber \\
 &  & \hspace*{4em}+\tilde{I}\tilde{J}\Big(4(45C_{2}{}^{(3)}+2\tilde{U})(u^{2}+v^{2})+9\tilde{J}(-1+3u^{2}+3v^{2})\Big)\Big]\nonumber \\
 &  & \quad+6C_{3}{}^{(2)}(C_{4}^{(3)}+2C_{5}^{(3)})\Big[-4(15C_{2}^{(3)}-2\tilde{J})\tilde{J}\tilde{U}(u^{2}+v^{2})+5\tilde{I}^{2}(3\tilde{J}+\tilde{U})(u^{2}+v^{2})\nonumber \\
 &  & \hspace*{4em}+\tilde{I}\Big(180C_{2}^{(3)}\tilde{J}(u^{2}+v^{2})+60C_{2}^{(3)}\tilde{U}(u^{2}+v^{2})\nonumber \\
 &  & \hspace*{6em}+18\tilde{J}^{2}(-1+3u^{2}+3v^{2})+\tilde{J}\tilde{U}(-6+13u^{2}+13v^{2})\Big)\Big]\Big\},
\end{eqnarray}
\begin{equation}
m_{2}^{(-1)}=\frac{9\tilde{I}\tilde{J}\tilde{U}}{4\tilde{\mathcal{J}}k^{2}u^{2}},
\end{equation}
\begin{eqnarray*}
m_{2}^{(1)} & = & -\frac{3}{4\tilde{\mathcal{J}}^{2}u^{2}}\Big\{2C_{3}^{(2)}\tilde{J}\Big[24C_{1}{}^{(3)}\tilde{J}\tilde{U}(-3+u^{2}+v^{2})\\
 &  & \hspace*{4em}+\tilde{I}^{2}\bigl(6\tilde{J}u^{2}-2\tilde{U}(u^{2}+2v^{2})\bigr)-\tilde{I}\tilde{U}\bigl(48(C_{1}^{(3)}+C_{2}^{(3)})(u^{2}+v^{2})+\tilde{J}(-3+2u^{2}+4v^{2})\bigr)\Big]\\
 &  & \quad+(C_{4}^{(3)}+2C_{5}^{(3)})\Big[2\tilde{I}^{2}u^{2}(9\tilde{J}^{2}-\tilde{U}^{2})\\
 &  & \hspace*{4em}+24\tilde{J}\tilde{U}\Big(2C_{2}^{(3)}\tilde{U}v^{2}+3C_{1}^{(3)}\tilde{J}(-3+u^{2}+v^{2})+C_{1}^{(3)}\tilde{U}(-3+u^{2}+3v^{2})\Big)\\
 &  & \hspace*{4em}-\tilde{I}\tilde{U}\Big(48(C_{1}^{(3)}+C_{2}^{(3)})u^{2}\tilde{U}+\tilde{J}(144(C_{1}^{(3)}+C_{2}^{(3)})u^{2}-3\tilde{U}+2u^{2}\tilde{U})+3\tilde{J}^{2}(-3+2u^{2}+4v^{2})\Big)\Big]\Big\},
\end{eqnarray*}
\begin{eqnarray}
m_{2}^{(3)} & = & \frac{3k^{2}}{2\tilde{\mathcal{J}}^{2}}\left(2C_{3}^{(2)}\tilde{J}+(C_{4}^{(3)}+2C_{5}^{(3)})(3\tilde{J}+\tilde{U})\right)\nonumber \\
 &  & \quad\times\Big\{\Big(24C_{2}^{(3)}C_{3}^{(2)}\tilde{I}+3(C_{4}^{(3)}+2C_{5}^{(3)})\tilde{I}\tilde{J}+2C_{3}^{(2)}\tilde{I}(\tilde{I}+\tilde{J})-12C_{2}^{(3)}(C_{4}^{(3)}+2C_{5}^{(3)})\tilde{U}\Big)(u^{2}+v^{2})\nonumber \\
 &  & \qquad+6C_{1}^{(3)}\Big[-2C_{3}^{(2)}\tilde{J}(-3+u^{2}+v^{2})+4C_{3}^{(2)}\tilde{I}(u^{2}+v^{2})\nonumber \\
 &  & \hspace*{4em}-3(C_{4}^{(3)}+2C_{5}^{(3)})\Big(\tilde{J}(-3+u^{2}+v^{2})+\tilde{U}(-1+u^{2}+v^{2})\Big)\Big]\Big\},
\end{eqnarray}
\begin{eqnarray}
m_{3}^{(0)} & = & -\frac{9\tilde{J}\tilde{U}^{2}}{8\tilde{\mathcal{J}}^{2}k^{2}u^{2}v^{2}}\Big\{6C_{1}^{(3)}\left[\tilde{J}(-3+u^{2}+v^{2})-2\tilde{I}(u^{2}+v^{2})\right]\nonumber \\
 &  & \qquad-\tilde{I}\left[\tilde{J}(-1+u^{2}+v^{2})+(12C_{2}^{(3)}+\tilde{I})(u^{2}+v^{2})\right]\Big\},
\end{eqnarray}
\begin{eqnarray}
m_{3}^{(2)} & = & \frac{3}{4\tilde{\mathcal{J}}^{2}u^{2}v^{2}}\Big\{2C_{3}^{(2)}\tilde{J}\Big[-6C_{1}^{(3)}\tilde{J}\tilde{U}(-3+u^{2}+v^{2})(u^{2}+v^{2})+\tilde{I}^{2}(u^{4}\tilde{U}-6\tilde{J}u^{2}v^{2}+\tilde{U}v^{4})\nonumber \\
 &  & \qquad+\tilde{I}\tilde{U}\Big(12C_{1}^{(3)}(u^{2}+v^{2})^{2}+12C_{2}^{(3)}(u^{2}+v^{2})^{2}+\tilde{J}(u^{4}+v^{4})\Big)\Big]\nonumber \\
 &  & \quad-3(C_{4}^{(3)}+2C_{5}^{(3)})\Big[(\tilde{I}\tilde{J}-4C_{2}^{(3)}\tilde{U})\bigl(2\tilde{I}u^{2}\tilde{U}v^{2}-\tilde{J}(u^{4}\tilde{U}-6\tilde{I}u^{2}v^{2}+\tilde{U}v^{4})\bigr)\nonumber \\
 &  & \qquad+2C_{1}^{(3)}\tilde{U}\Bigl(-4\tilde{I}u^{2}\tilde{U}v^{2}+3\tilde{J}^{2}\bigl(u^{4}+v^{2}(-3+v^{2})+u^{2}(-3+2v^{2})\bigr)\nonumber \\
 &  & \hspace*{4em}+\tilde{J}\bigl(3u^{4}\tilde{U}+3\tilde{U}v^{2}(-1+v^{2})+u^{2}(-3\tilde{U}-12\tilde{I}v^{2}+2\tilde{U}v^{2})\bigr)\Bigr)\Big]\Big\},
\end{eqnarray}
\begin{equation}
m_{4}^{(0)}=\frac{9\tilde{I}\tilde{J}\tilde{U}}{8\tilde{\mathcal{J}}k^{2}u^{2}},
\end{equation}
\begin{eqnarray}
m_{4}^{(2)} & = & -\frac{1}{4\tilde{\mathcal{J}}^{2}u^{2}}\Big\{9(C_{4}^{(3)}+2C_{5}^{(3)})\Big[2\tilde{J}\tilde{U}\Big(2C_{2}^{(3)}\tilde{U}v^{2}+3C_{1}^{(3)}\tilde{J}(-3+u^{2}+v^{2})+C_{1}^{(3)}\tilde{U}(-3+u^{2}+3v^{2})\Big)\nonumber \\
 &  & \hspace*{4em}+\tilde{I}^{2}\tilde{J}u^{2}(3\tilde{J}+\tilde{U})-\tilde{I}\tilde{U}\Big(4C_{1}^{(3)}u^{2}(3\tilde{J}+\tilde{U})+4C_{2}^{(3)}u^{2}(3\tilde{J}+\tilde{U})+\tilde{J}^{2}v^{2}\Big)\Big]\nonumber \\
 &  & \quad+6C_{3}^{(2)}\tilde{J}\Big[6C_{1}^{(3)}\tilde{J}\tilde{U}(-3+u^{2}+v^{2})+\tilde{I}^{2}(3\tilde{J}u^{2}-\tilde{U}v^{2})\nonumber \\
 &  & \hspace*{4em}-\tilde{I}\tilde{U}\Big(\tilde{J}v^{2}+12C_{1}^{(3)}(u^{2}+v^{2})+12C_{2}^{(3)}(u^{2}+v^{2})\Big)\Big]\Big\},
\end{eqnarray}
\begin{equation}
m_{5}^{(1)}=-\frac{9\tilde{J}\tilde{U}}{8\tilde{\mathcal{J}}^{2}k^{2}u^{2}v^{2}}\left[2C_{1}^{(3)}\tilde{J}\tilde{U}(-3+u^{2}+v^{2})+\tilde{I}^{2}\tilde{J}(u^{2}+v^{2})-4(C_{1}^{(3)}+C_{2}^{(3)})\tilde{I}\tilde{U}(u^{2}+v^{2})\right].
\end{equation}
where we define the combinations of coefficients that are constant
with the power-law solution ansatz
\begin{align*}
\tilde{U} & =\frac{a^{2}}{\mathcal{H}^{2}}U,I=\frac{a^{2}}{\mathcal{H}^{2}}I,\tilde{K}=\frac{a^{2}}{\mathcal{H}^{2}}K,\tilde{J}=\frac{a^{2}}{\mathcal{H}^{2}}J,\tilde{\mathcal{J}}==\frac{a^{4}}{\mathcal{H}^{4}}\mathcal{J}.
\end{align*}

Various coefficients in the the source function $f$ in (\ref{srcf})
are given by
\begin{align*}
d_{0}^{-} & =\frac{m_{1}^{(2)}}{4k^{2}uv}-\frac{m_{2}^{(3)}}{2k^{2}uv}+\frac{1}{4}m_{3}^{(2)}-\frac{vm_{4}^{(2)}}{2u}+(u\leftrightarrow v),\\
d_{2}^{-} & =\frac{1}{4}k^{2}m_{3}^{(0)}-\frac{k^{2}vm_{4}^{(0)}}{2u}+\frac{k^{2}vm_{5}^{(1)}}{2u}-k^{2}m_{5}^{(1)}+\frac{m_{1}^{(0)}}{4uv}-\frac{m_{2}^{(1)}}{2uv}+\frac{m_{3}^{(2)}}{4uv}+\frac{m_{4}^{(2)}}{uv}+(u\leftrightarrow v),\\
d_{4}^{-} & =-\frac{k^{2}m_{2}^{(-1)}}{2uv}+\frac{k^{2}m_{3}^{(0)}}{4uv}+\frac{k^{2}m_{4}^{(0)}}{uv}-\frac{k^{2}m_{5}^{(1)}}{uv}+(u\leftrightarrow v),\\
d_{x}^{-} & =\frac{m_{2}^{(3)}}{2k^{2}u}-(u\leftrightarrow v),\\
d_{1}^{-} & =\frac{1}{2}k^{2}vm_{5}^{(1)}+\frac{m_{2}^{(1)}}{2u}+\frac{m_{3}^{(2)}}{4v}-\frac{m_{3}^{(2)}}{4u}+\frac{m_{4}^{(2)}}{v}-(u\leftrightarrow v),\\
d_{3}^{-} & =\frac{k^{2}m_{2}^{(-1)}}{2u}+\frac{k^{2}m_{3}^{(0)}}{4v}-\frac{k^{2}m_{3}^{(0)}}{4u}-\frac{k^{2}m_{4}^{(0)}}{u}+\frac{k^{2}m_{5}^{(1)}}{u}-\frac{k^{2}m_{5}^{(1)}}{v}-(u\leftrightarrow v).
\end{align*}
and $d_{i}^{-}(u,v)=d_{i}^{+}(u,-v)$.

\providecommand{\href}[2]{#2}\begingroup\raggedright\endgroup

\end{document}